\documentclass[reprint,aps,pre,twocolumn,notitlepage,superscriptaddress,longbibliography]{revtex4-1}
\usepackage[normalem]{ulem}
\usepackage{float}
\usepackage{graphicx,amsmath,amssymb}
\usepackage{epsfig,comment}
\usepackage{epstopdf}
\usepackage{dcolumn}
\usepackage{bm}
\usepackage{bm,color}
\usepackage{float}
\definecolor{green}{RGB}{0,128,0}

\newcommand{\nn}{\langle\boldsymbol{nn}\rangle}

\begin{document}

\title{
Constitutive model for shear-thickening suspensions: \\ Predictions for steady shear with superposed transverse oscillations}
\author{J. J. J. Gillissen$^1$, C. Ness$^2$, J. D. Peterson$^3$, H. J. Wilson$^1$ and M. E. Cates$^3$} 
\affiliation{$^1$ Department of Mathematics, University College London, Gower Street, London WC1E 6BT, United Kingdom\\
$^2$ Department of Chemical Engineering and Biotechnology, University of Cambridge,  Cambridge CB3 0AS, United Kingdom\\
$^3$DAMTP, Centre for Mathematical Sciences, University of Cambridge, Cambridge CB3 0WA, United Kingdom}

\date{\today}
\begin{abstract}
We recently developed a tensorial constitutive model for dense, 
shear-thickening particle suspensions that combines rate-independent microstructural evolution with a stress-dependent jamming threshold. This gives a good qualitative account for reversing flows, although it quantitatively over-estimates structural anisotropy  [J.~J.~J.~Gillissen {\em et al.}, Phys. Rev. Lett. {\bf 123} (21), 214504 (2019)]. Here we use the model to predict the unjamming effect of superposed transverse oscillations on a steady shear flow in the thickened regime [N.~Y.~C.~Lin {\em et al.}, Proc.~Nat.~Acad.~Sci.~USA {\bf 113}, 10774 (2016)].
The model successfully reproduces the oscillation-mediated viscosity drop observed experimentally.
We compare the time-dependent components of the stress and microstructure tensors to discrete-element simulations.
Although the model correctly captures the 
main qualitative behaviour, it generally over-predicts the 
microstructural anisotropy in steady shear, and it under-predicts the number of particle contacts 
in oscillating shear. It also does not fully capture the correct variation in phase angle between the transverse component of the microstructure and the shear rate oscillations, as the amplitude of the latter is increased. These discrepancies suggest avenues for future improvements to the model.
\end{abstract}
\maketitle

\section{Introduction}	
Dense suspensions of hard particles in a viscous solvent are found in many application domains including the construction industry, food production, and pharmaceuticals.
Such materials, which have solid volume fraction $\phi\gtrsim0.4$, often exhibit shear thickening, 
an increase (continuous or discontinuous) in viscosity under increasing shear rate $\dot{\gamma}$. Understanding and controlling this distinctive rheological behaviour is key to operating efficient and reliable processes, and has been the subject of much study during the past three decades.

Recent numerical~\cite{mari2014shear,seto2013discontinuous} and experimental~\cite{guy2015towards,lin2015hydrodynamic,royer2016rheological} data provide evidence that, in contrast to scenarios envisaged in much of the prior literature~\cite{wagner2009shear}, shear thickening in non-Brownian, non-inertial suspensions is caused by the onset of direct interparticle contacts that are frictional in character.
In addition to tangential contact friction forces, shear thickening may also arise due to tangential lubrication forces, 
that act between asperities on the opposing particle surfaces \cite{jamali2019alternative}. In both (the contact friction and the lubrication) scenarios, shear thickening results from constraints due to tangential forces. Although our theoretical treatment of both scenarios would be similar, we follow the contact friction narrative in this work. 
Experimental data~\cite{guy2015towards} for the steady-state viscosity as a function of shear rate are well described by the theory of Wyart and Cates (WC)~\cite{wyart2014discontinuous} in which the appearance of such contacts under steady flow is governed by a competition between
a short-ranged interparticle repulsion, of maximum force $F^*$,
and the macroscopic particle pressure $\Pi = -\mathrm{Tr}\boldsymbol{\Sigma}/3$, 
with $\boldsymbol\Sigma$ the particle stress tensor.
In suspensions of strictly hard spheres, whether frictional or not, dimensional analysis predicts rate-independent rheology, {\em i.e.}, $\boldsymbol{\Sigma}\propto\dot{\gamma}$~\cite{boyer2011unifying}. However, the presence of a characteristic force scale $F^*$ allows the physics to depend on a
dimensionless shear rate: 
\begin{equation}
\dot{\gamma}_r = \frac{\dot{\gamma}\eta_s}{\Pi^*},
\label{eq10}
\end{equation}
with $\eta_s$ the solvent viscosity, $a$ the particle radius, and $\Pi^*\sim F^*/a^2$ the so-called `onset stress'.
At small flow rates, where $\Pi< \Pi^*$, the typical interparticle force remains less than $F^*$, and particles remain separated by lubrication films \cite{comtet2017pairwise}.
At large flow rates,  where $\Pi> \Pi^*$,
lubrication films break down
and particles enter into solid-solid frictional contact. 
Friction restricts particle sliding
so that steady flow requires more tortuous particle trajectories,
leading to an increase in the suspension viscosity.

Based upon this principle, WC write, for a steady $xy$ shear flow, a relation between the nondimensionalized suspension viscosity $\eta_r=\Sigma_{xy}/(\eta_s\dot{\gamma})$, the volume fraction
$\phi$ and the dimensionless shear rate $\dot{\gamma}_r$.
The relation is based on the Krieger--Dougherty equation \cite{krieger1959mechanism}:
\begin{equation}
\eta_r \sim (1-\phi/\phi^J)^{-2},
\label{eq402}    
\end{equation}
where $\phi^J$ is the volume fraction at jamming. 
WC introduce rate-dependence by relating $\phi^J$ to the onset of friction described above, 
noting that friction imposes additional constraints at particle contact, 
reducing the number of contacts per particle (or coordination number, $Z$) required for jamming.
Moreover they effectively assume that the steady-state microstructure itself is friction-independent, so that the stress dependence enters not by changes in $Z$ itself, but by changes in the jamming point $Z^J$, which for spheres in three dimensions can vary between $4$ (all contacts rolling) and $6$ (all contacts sliding). 
This assumption causes the steady-state $Z$ value to depend solely on volume fraction, so
reducing $Z^J$ is equivalent to reducing $\phi^J$.
Hence WC postulated:
\begin{equation}
\phi^J(f)=\phi^J_1(1-f)+\phi^J_2 f\quad,\quad f(\Pi)= \exp\left(-{\Pi^*}/{\Pi}\right)\text{,}
\label{eq401}
\end{equation}
where $f(\Pi)$ is the fraction of contacts that are constrained by friction to roll rather than slide.
(The particular form of $f(\Pi)$ is relatively unimportant; the above choice was made later, on empirical grounds, in Ref. \cite{hermes2016unsteady}.)
The limiting volume fractions at which frictionless and fully frictional packings become rigid,
$\phi^J_1\approx0.64$ and $\phi^J_2\approx 0.57$ respectively (in 3D),
are generally agreed upon in the literature. 

Although Eqs.~(\ref{eq402}, \ref{eq401})
have a featureless, monotonic dependence of $f$ on $\Pi$,
they predict flow curves (shear stress versus shear rate) that, depending on $\phi$,
imply continuous and discontinuous shear thickening as well as `full jamming' (whereby the viscosity is infinite above a $\Pi$ threshold comparable in magnitude to $\Pi^*$). In particular, discontinuous shear thickening arises as a jump between the lower and upper branches of a flow curve that is everywhere smooth, but $S$-shaped~\cite{wyart2014discontinuous}.

The WC theory agrees well with experiments and particle-based simulations under steady and homogeneous conditions~\cite{guy2015towards}, at least for modest particle size polydispersity~\cite{guy2020testing}. Its predictions of non-monotonic flow curves also signal the presence of steady shear-banding, and other instabilities leading to spatiotemporal variations of the flow state \cite{hermes2016unsteady}. However, it makes no predictions for unsteady flow, nor does it quantitatively address the tensorial character of the stress tensor. In other words, WC did not offer a full constitutive model for shear-thickening suspensions.
At first sight one might consider applying the WC equations (\ref{eq402}, \ref{eq401}) at each point in time during an evolving flow, but the resulting implicit assumption that the coordination number $Z$ depends only upon $\phi$ is clearly invalidated by the flow-history dependence of the microstructure.

To address this, we have recently formulated a tensorial
constitutive model in which the viscosity depends on a time-evolving `jamming coordinate' $\xi$, defined in Eq. (\ref{eq21}) below, 
which can take over the role played by $\phi$ in the WC theory ~\cite{gillissen2019constitutive}.
Although $\xi$ is effectively a proxy for a time-evolving microscopic coordination number $Z$,  the jamming coordinate is computable from the {\em coarse-grained} microstructure 
$\langle\boldsymbol{nn}\rangle$ [see Eq. (\ref{eq15}) below], 
allowing closure of our equations at that level. 
Our model marries a microstructure-tensor evolution equation, which was
derived previously for rate-independent suspensions~\cite{gillissen2018modeling}
from which $\xi$ is computed,
with key intuitions for shear-thickening suspensions as described in the scalar and time-independent WC approach~\cite{wyart2014discontinuous}.
These are
the (linear) interpolation between jamming conditions as a function of $f(\Pi)$
and the singular  (Krieger-Dougherty) dependence of viscosity on $1-\xi/\xi^J$, 
where $\xi^J$ is the jamming coordinate at the jamming point which is defined in Eq. (\ref{eq2}) below.
In Ref.~\cite{gillissen2019constitutive}, 
we demonstrated that the new constitutive model 
performs well under shear reversal,
correctly predicting
the discontinuous drop in $\eta_r = \Sigma_{xy}/(\dot{\gamma}\eta_s)$ at very small strain and its subsequent smooth recovery.
Abrupt flow reversals of this kind represent important test cases,
which in the literature have been used 
to gain insight into history-dependent microstructure~\cite{Gadala-maria1980}
and to distinguish the contact and hydrodynamic contributions to suspension stress~\cite{lin2015hydrodynamic,Ness2016b,peters_rheology_2016}.
Their challenging character for constitutive models has been previously pointed out~\cite{goddard2006dissipative,chacko2018shear}.

\begin{figure}
\centerline{
\includegraphics[trim={0mm 0mm 0mm 0},clip,width = 1\linewidth]{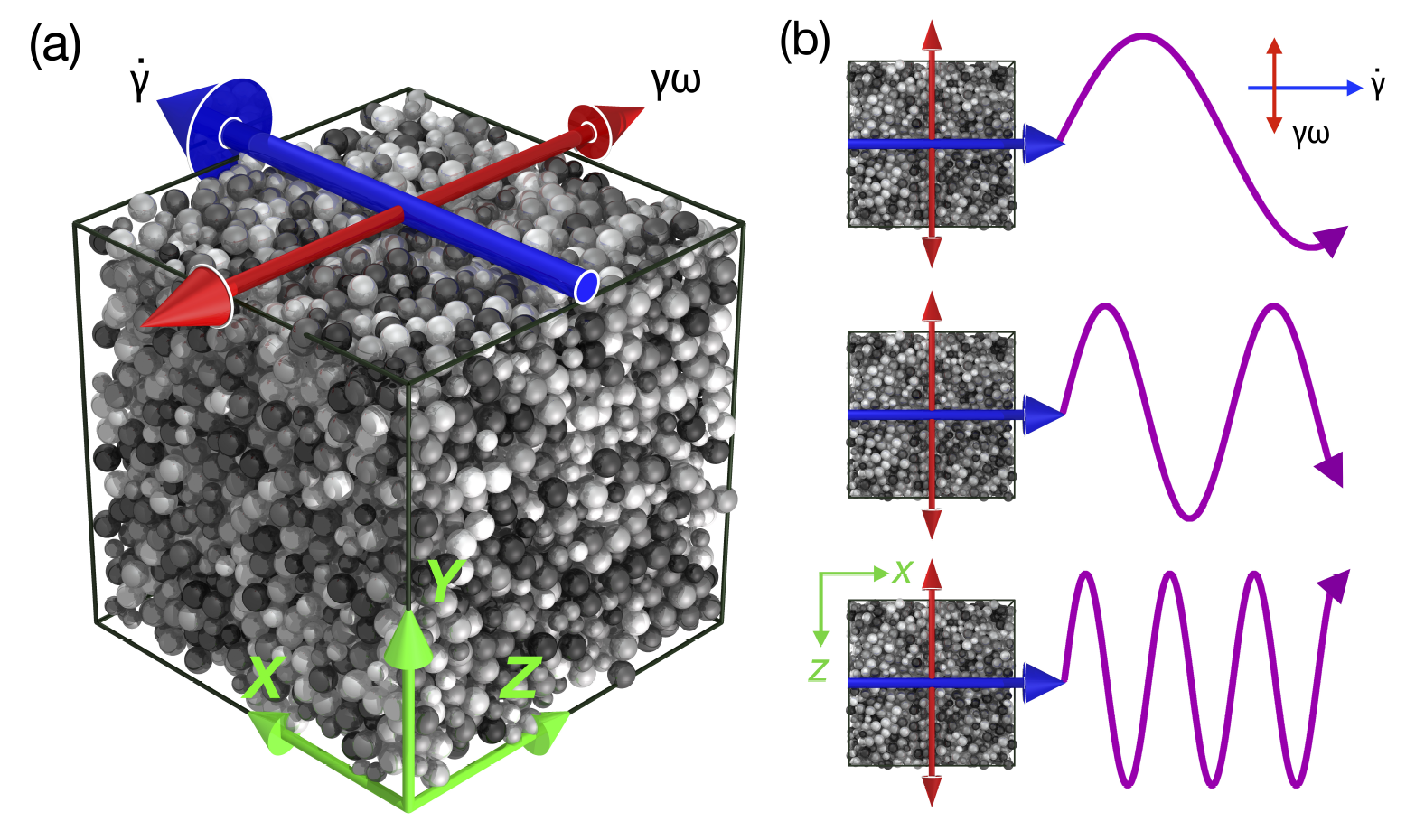}
}
\caption{
(a) Model suspension with coordinate definition in green. Blue arrow indicates steady shear flow; red arrows indicate superposed transverse oscillatory shear flow.
(b) Top view of model suspension showing examples of shearing trajectories with different $\dot{\gamma}_\perp = \gamma\omega/\dot{\gamma}$.
}
\label{figure1}
\end{figure}

In the present work we further test the new constitutive model by addressing the case of a steady shear flow (shear rate $\dot\gamma$) with superposed transverse shear flow oscillations. 
The latter is at $90^\circ$ to the steady flow and has frequency $\omega$ and strain amplitude $\gamma$, Fig.~\ref{figure1}a. 
The steady shear has flow in $x$ and gradient in $y$, 
while the transverse oscillations are in $\pm z$ with gradient in $y$.
As opposed to an abrupt flow reversal, such flows constitute a continuous family of time-dependent,
controlled distortions to the steady shear flow,
characterized by their amplitude $\gamma$ (which, as discussed later, we fix at $\gamma= 0.01$ following the experiments of Ref.~\cite{lin2016tunable}) and a dimensionless frequency $\dot{\gamma}_\perp$; see Fig.~\ref{figure1}b:
\begin{equation}
\dot{\gamma}_\perp = \frac{\omega\gamma}{\dot{\gamma}}.
\label{eq40}
\end{equation}

Recent experiments~\cite{lin2016tunable} and DEM simulations~\cite{ness2018shaken} demonstrate 
that the oscillations
break up the fragile jammed network of interparticle contacts ~\cite{cates1998jamming}. This break-up can
substantially reduce the viscosity in systems with $\phi$ just below $\phi^J_2$ (where discontinuous shear thickening arises). Moreover, for systems that are fully jammed ($\phi>\phi^J_2, \Pi\gg\Pi^*$) the viscosity falls from the (effectively infinite) jammed value to a finite one.  
The mechanism behind the loss of contacts has been explained
from the perspectives of force chain dynamics~\cite{lin2016tunable}
and random organisation~\cite{ness2018shaken}.
Transverse shear flow oscillations may enhance suspension flow in practical applications~\cite{sehgal2019using}.
In addition, this flow configuration also offers a subtle and challenging test case for
constitutive models for suspension microstructure and stress. 

The remainder of the paper is organised as follows.
In Sec.~\ref{continuum} we give a self-contained derivation of our constitutive model, including those parts first presented in 
\cite{gillissen2018modeling,gillissen2019effect} as well as the new features added in \cite{gillissen2019constitutive} to address shear thickening.
In Sec.~\ref{particle} we give brief details of the discrete-element simulation model from which we generate test data in the chosen flow geometry.
In Sec.~\ref{results} we compare the results of the constitutive model to
those of the discrete-element model across a range of $\dot{\gamma}_r$ and $\dot{\gamma}_\perp$. 
Our conclusions are given in Sec.~\ref{conclusions}.

\section{Constitutive Model}
\label{continuum}
We consider a collection of non-Brownian spheres of radius $a$, suspended 
at a volume fraction of $\phi$ and a number density of $n=\phi/(\tfrac{4}{3}\pi a^3)$ in a fluid of 
density $\rho$ and viscosity $\eta_s$. The 
volume-averaged fluid velocity is $\boldsymbol{U}$, 
and the fluid velocity gradient and deformation tensors are given by
$\boldsymbol{L}=\boldsymbol{\nabla U}^T$
and
$\boldsymbol{E}=\tfrac{1}{2}\left(\boldsymbol{L}+\boldsymbol{L}^T\right)$,
respectively. 
The particle Reynolds number is $\dot{\gamma}\rho a^2/\eta_s\ll 1$ (allowing inertia to be neglected)
and $\phi$ is assumed sufficiently large that hydrodynamic interactions between particles can effectively be modelled as lubrication forces. 

Below we derive an equation for the particle stress tensor $\boldsymbol{\Sigma}$ which is based on 
an equation of motion for the statistics of the particle pair separation unit vector $\boldsymbol{n}$, 
which is encoded in the 
second moment $\langle\boldsymbol{nn}\rangle$ of the distribution function $\Psi(\boldsymbol{n})$.
In Sec. \ref{pairmotion} we derive an equation of motion for $\boldsymbol{n}$ for a single particle pair.
In Sec. \ref{micro} we use this equation to derive the equation of motion for $\langle\boldsymbol{nn}\rangle$ and
in Sec. \ref{stress} we relate $\boldsymbol{\Sigma}$ to $\langle\boldsymbol{nn}\rangle$.

\subsection{Particle Pair Motion}
\label{pairmotion}
Following \cite{gillissen2018modeling}, we start by writing an equation of motion for the connection vector $\boldsymbol{r}$ of a particle pair, 
that points to
a so-called `test particle' (TP) from a so-called `pairing particle' (PP).
Under the conditions given above, Newton's equation of motion, applied to the TP reads:
\begin{multline}
\boldsymbol{0}=
C_1a\eta_s\left(\boldsymbol{L\cdot r}-\dot{\boldsymbol{r}}\right)\\
-C_2a\eta_s\left(\dot{\boldsymbol{r}}\cdot \boldsymbol{n}\right)\boldsymbol{n}
+C_3a^2\eta_s\dot{\gamma}\Theta\left(2a-r\right)\boldsymbol{n}.
\label{eq1}    
\end{multline}
Here $\boldsymbol{n}=\boldsymbol{r}/r$ is the interaction unit vector, 
$r=\vert\boldsymbol{r}\vert$, 
$C_{1,2,3}$ are dimensionless pre-factors, specified below,
and 
$\Theta(u)$ is the Heaviside step-function, with $\Theta(u<0) = 0$, $\Theta (u\ge 0) = 1$. 
For strictly hard-core particles the $\Theta$ function counts contacts; in systems where hard-core contact is replaced by particle overlaps (as is often done in simulations) it continues to do so. Note that in Ref.~\cite{gillissen2018modeling} we wrote $C_1/2$ instead of $C_1$.

The $C_1$-term in Eq.~(\ref{eq1}) 
is the interaction force between the TP and the background mixture, which
is proportional to the difference between 
the TP velocity $\dot{\boldsymbol{r}}$ and the 
mixture velocity at the TP location, $\boldsymbol{L\cdot r}$. 
The $C_2$-term in Eq. (\ref{eq1}) is 
the lubrication interaction force 
between the TP and the PP. 
The leading order contribution to the lubrication force $\sim a^2(r-2a)^{-1}\eta_s \left(\dot{\boldsymbol{r}}\cdot \boldsymbol{n}\right)\boldsymbol{n}$, where $(r-2a)$ is the interparticle gap \cite{kim1991microhydrodynamics}. 
In order to arrive at tractable expressions for the suspension microstructure and stress [Eqs. (\ref{eq15}, \ref{eq19})], 
we have replaced the factor $a(r-2a)^{-1}$ with its averaged value $C_2$, 
which is taken to obey the Krieger-Dougherty form \cite{krieger1959mechanism}: 
\begin{equation}
C_2\sim(1-\phi/\phi^J_1)^{-2},
\label{eq1b}
\end{equation}
where $\phi_1^J$ is the particle volume fraction at random close packing. 

The $C_3$-term in Eq. (\ref{eq1}) is the contact force 
between the TP and the PP. 
The expression for the contact force assumes that this force 
(i) aligns with $\boldsymbol{n}$, 
(ii) acts on the particle surface, 
(iii) scales as a viscous force $\sim a^2\eta_s\dot{\gamma}$ and 
(iv) is proportional to a dimensionless pre-factor $C_3$. 

Note that in treating $C_{1,2,3}$ as constants, independent of local
microstructure, we have already used a mean-field type of averaging. (This applies particularly for the constraint force 
$C_3$ which, at a particular contact, can take any positive value to balance the other forces acting.)
After such averaging, the interaction force between the TP and the background must balance 
the dominant term of the interaction force with the PP, so that in magnitude 
\begin{equation}
C_1\sim\max(C_2,C_3).
\label{eq1a}
\end{equation}

Note that there are no tangential (lubrication or contact friction) forces in Eq.~\eqref{eq1} and we do not consider the torque balance.  
When considering particle motion, omission of tangential contact forces, caused by friction, is justified by the assumption, inherited from the WC theory, that microstructural evolution is not itself altered by frictional forces (although the stress for a given microstructure and flow is strongly altered). 
This assumption is further justified by observations from DEM (i) that the magnitude of the tangential contact forces 
is small compared to that of the normal contact forces, even under shear-thickened conditions \cite{seto2018normal},
and (ii) that the microstructure is nearly unaffected by shear thickening \cite{gillissen2019constitutive}.
Indirectly the tangential contact friction forces \textit{are} important as they affect the suspension rheology by imposing constraints on the particle motion \cite{wyart2014discontinuous}.
The resulting increase in the suspension viscosity is, however, mainly supported by the normal contact forces.  
Therefore, although we exclude the tangential contact friction forces in Eq. (\ref{eq1}), 
we indirectly account for these forces by incorporating 
the following jamming behaviour in the pre-factor $C_3$ for the normal contact forces:
\begin{equation}
C_3\sim (1-\xi/\xi^J)^{-2},
\label{eq1c}
\end{equation}
which depends on the jamming coordinate $\xi$, a mesoscopic quantity, defined in Eq. (\ref{eq21}) below. 
The jamming coordinate $\xi$ serves as a proxy for the coordination number $Z$, as $\phi$ does in the steady-state WC theory. 
Although the numerical values of $\xi$ and $Z$ differ, as exemplified in Eq. (\ref{eq33}) below,
$\xi$ plays a similar role as $Z$, 
by defining a distance to the jamming point, 
i.e. $C_3$ diverges when $\xi$ reaches its jamming limit $\xi^J$ [Eq. (\ref{eq1c})]. 
This jamming limit $\xi^J$ is in turn assumed to 
decrease from a larger value $\xi^J_1$ to a smaller value $\xi_2^J$, when the 
system transitions from `lubricated' to `frictional'. 
This transition is encoded in the fraction $f$ of frictional contacts, 
which smoothly increases from zero to one as the particle
pressure in the system, $\Pi$, 
passes through the onset threshold $\Pi^*\sim F^*/a^2$. Here $F^*$ is the maximum force sustainable by the short-range repulsive interactions:
\begin{equation}
\xi^J=\xi_1^J\left(1-f\right)+\xi_2^J f,\hspace{1cm} f=\exp\left(-{\Pi^*}/{\Pi}\right).
\label{eq2}
\end{equation}
For simplicity we have adopted the same functional form for $f(\Pi)$ as in Eq.~(\ref{eq401}).

The normal and tangential components of the inter-particle velocity $\dot{\boldsymbol{r}}$ are readily obtained by projecting Eq.~(\ref{eq1}) onto the relevant directions:
\begin{equation}
\dot{\boldsymbol{r}}\cdot\boldsymbol{n}=\frac{C_1}{C_1+C_2}\boldsymbol{E:rn}+\frac{C_3}{C_1+C_2}\dot{\gamma}a\Theta\left(2a-r\right),
\label{eq4}
\end{equation}
and, with $\boldsymbol{\delta}$ the unit tensor,
\begin{equation}
\dot{\boldsymbol{r}}\cdot\left(\boldsymbol{\delta}-\boldsymbol{nn}\right)=\boldsymbol{L}\cdot\boldsymbol{r}\cdot\left(\boldsymbol{\delta}-\boldsymbol{nn}\right).
\label{eq5}
\end{equation}

\subsection{Microstructure Evolution}
\label{micro}
Again following \cite{gillissen2018modeling}, we now introduce the distribution function $\Psi\left(\boldsymbol{r}\right)$ of the particle-pair separation vector $\boldsymbol{r}$, which evolves according to the Smoluchowski equation 
for the two-particle configuration space:
\begin{equation}
\partial_t\Psi+\partial_{k}\left(\dot{r}_k\Psi\right)=0,
\label{eq6}
\end{equation}
where $\partial_{k}=\partial/\partial r_k$.
Because the typical spacing between the particles $\epsilon$ is small 
compared to the particle radius $a$, the anisotropy in $\Psi(\boldsymbol{r})$ 
is relegated to the so-called coarse-graining shell $2a<r<2a+\epsilon$,
where $\epsilon$ is related to $\phi$ via $[a/(a+\epsilon)]^3\sim\phi/\phi^J_1$, i.e.:
\begin{equation}
\frac{\epsilon}{a}\sim1-\phi/\phi^J_1.
\label{eq777}
\end{equation}
By assuming that the number of interactions in the coarse-graining shell $\sim\phi$, 
we see that 
$\Psi(\boldsymbol{r})\sim [\mathrm{number\;of\;interactions}]/[\mathrm{volume\;of\;shell}]\sim{\phi}/({a^2\epsilon})$.
Outside the coarse-graining shell, steric constraints are dominant and 
$\Psi(r=[2a+\epsilon]^{+})=\Psi^{\mathrm{outer}}$ is assumed isotropic. 
By continuity we write this as: 	
\begin{equation}
\Psi^{\mathrm{outer}}= \frac{\phi}{4\pi a^2\epsilon}.
\label{eq666}
\end{equation}
Eq. (\ref{eq666}) suppresses an order-unity
pre-factor that can, however, be absorbed into other constants appearing below.


Next, we derive the evolution equation for the 
second-order orientation moments $\langle\boldsymbol{nn}\rangle$ of the 
distribution function $\Psi(\boldsymbol{r})$ in the coarse-graining shell, by 
inserting Eqs.~(\ref{eq4}, \ref{eq5}) into Eq.~(\ref{eq6}), 
multiplying the result with $\boldsymbol{nn}$, 
applying the following, so-called coarse-graining operator $\langle\cdots\rangle$: 
`
and approximating $\boldsymbol{r}\approx 2a\boldsymbol{n}$ \cite{gillissen2018modeling}:
\begin{multline}
\partial_t\langle \boldsymbol{nn}\rangle ={\boldsymbol{L}}\cdot\langle \boldsymbol{nn}\rangle+\langle \boldsymbol{nn}\rangle\cdot{\boldsymbol{L}}^T-
2{\boldsymbol{L}}:\langle \boldsymbol{nnnn}\rangle\\
-\frac{2aC_1}{C_1+{C}_2}{\boldsymbol{E}}:\oint_{r=2a+\epsilon} \Psi(\boldsymbol{r}) \boldsymbol{nnnn}\, d^2\boldsymbol{r}.
\label{eq8}
\end{multline}
The boundary surface integral in Eq.~(\ref{eq8})
corresponds to an orientation probability flux 
between the coarse-graining shell $2a<r<2a+\epsilon$ and the outer shell $r>2a+\epsilon$.
This flux is carried by the rate of strain tensor: 
${\boldsymbol{E}}=\boldsymbol{E}_c+\boldsymbol{E}_e$, 
which is decomposed into its positive and negative eigen-parts.
For instance in simple $xy$-shear flow: 
\begin{equation}
\boldsymbol{E}_e=\tfrac{1}{2}\dot{\gamma}\boldsymbol{n}_e\boldsymbol{n}_e,\hspace{1cm}
\boldsymbol{E}_c=-\tfrac{1}{2}\dot{\gamma}\boldsymbol{n}_c\boldsymbol{n}_c, 
\label{eq11}
\end{equation}
where $\pm \tfrac{1}{2}\dot{\gamma}$ are the expansive and the compressive eigenvalues of $\boldsymbol{E}$ and 
$\boldsymbol{n}_e=(1,1,0)/\sqrt{2}$ and $\boldsymbol{n}_c=(1,-1,0)/\sqrt{2}$ are the corresponding eigenvectors.

The positive (extensional) eigen-part $\boldsymbol{E}_e$ and negative (compressive) eigen-part $\boldsymbol{E}_c$
correspond to an outward and an inward probability flux
between the coarse-graining shell and the outer shell, respectively. 
Note that the contact force 
[$C_3$-term in Eq. (\ref{eq1})] 
does not enter Eq.~\eqref{eq8}; this reflects the fact that, for impenetrable particles, there is no probability flux across the inner surface of the coarse-graining shell at $r=2a$. Consequently, within our model the evolution of the coarse-grained 
microstructure tensor $\langle \boldsymbol{nn}\rangle$ is not directly sensitive to contact forces.


With these assumptions 
the surface integral in Eq.~(\ref{eq8}) can now be recast as \cite{gillissen2018modeling}:
\begin{multline}
{\boldsymbol{E}}:\oint_{r=2a+\epsilon} \Psi(\boldsymbol{r}) \boldsymbol{nnnn} \,d^2\boldsymbol{r}=\\
\epsilon^{-1}{\boldsymbol{E}_e}:\langle\boldsymbol{nnnn}\rangle+
\epsilon^{-1}{\boldsymbol{E}_c}:\langle\boldsymbol{nnnn}\rangle^{\mathrm{outer}},
\label{eq9}
\end{multline}
where 
$\langle\boldsymbol{nnnn}\rangle$ and $\langle\boldsymbol{nnnn}\rangle^{\mathrm{outer}}$ are 
the orientation moments, evaluated inside the coarse-graining shell and on the outside of the coarse-graining shell [Eq. (\ref{eq666})], 
respectively.
Combining Eqs.~(\ref{eq8}, \ref{eq9}), 
we obtain the following coarse-grained microstructure evolution equation:
 \begin{multline}
\partial_t\langle \boldsymbol{nn}\rangle =
{\boldsymbol{L}}\cdot\langle \boldsymbol{nn}\rangle+\langle \boldsymbol{nn}\rangle\cdot{\boldsymbol{L}}^T-
2{\boldsymbol{L}}:\langle \boldsymbol{nnnn}\rangle\\
-\beta\left[
\boldsymbol{E}_e: \langle \boldsymbol{nnnn}\rangle+
\boldsymbol{E}_c: \langle \boldsymbol{nnnn}\rangle^{\mathrm{outer}}
 \right],
  \label{eq12}
\end{multline}
where $\beta$ is referred to as the microstructure association rate:
\begin{equation}
\beta=\frac{2aC_1}{\epsilon\left(C_1+{C}_2\right)},
\label{eq12c}
\end{equation} 
which controls the rate of particle pair association and dissociation. 
The physical importance of $\beta$ is that its inverse sets a strain scale for structural evolution. 
On geometrical grounds, $\beta$ 
should depend on $\phi$ so as to diverge at random close packing $\phi_1^J$. 
To determine the dependence of $\beta$ on $\phi$, 
we 
make use of Eq. (\ref{eq777}).
We furthermore see from [Eq.~(\ref{eq1a})], that 
$C_1\gtrsim C_2$, such that $C_1+C_2\sim C_1$.
Inserting these approximations in Eq. (\ref{eq12c}),
we find:
\begin{equation}
\beta=\frac{\beta_0}{1-{\phi}/{\phi^J_1}},
\label{eq999}
\end{equation} 
with $\beta_0$ a tuneable parameter.
This shows that $\beta$ 
is roughly constant in the region just below $\phi^J_2$ where shear thickening is seen.
In Ref. \cite{gillissen2019constitutive} 
we determine $\beta$ by matching in this region the constitutive model 
to DEM simulation data after reversal of steady shear. 

The first line of Eq.~(\ref{eq12})  describes the rotational advection
of the contact vectors $\boldsymbol{n}$,
whereas the second line corresponds to the association and dissociation of interacting particle pairs 
by the action of compressive and extensional flow deformations
that, respectively, push particles together and pull them apart. More specifically, the compressive rate of strain $\boldsymbol{E}_c$ advects, into
the coarse-graining shell, the isotropic exterior distribution of non-contacting particles, importing preferentially along the compression
axis (or axes). In contrast the extensional rate of strain $\boldsymbol{E}_e$
advects the anisotropically distributed existing contacts
out of the coarse-graining shell, exporting preferentially
along the extension axis (or axes).

Eq.~(\ref{eq12}) contains the fourth order moment of the 
inner probability density function $\langle\boldsymbol{nnnn}\rangle$ and 
of the isotropic outer distribution function $\langle\boldsymbol{nnnn}\rangle^{\mathrm{outer}}$.
We next express $\langle\boldsymbol{nnnn}\rangle$ in terms of
$\langle\boldsymbol{nn}\rangle$,
using the linear closure model of Hinch and Leal, 
which is accurate for microstructures that are relatively close to isotropy \cite{hinch1976constitutive}:
\begin{multline}
\langle n_in_jn_kn_l\rangle=
-\frac{1}{35}
\langle n_mn_m\rangle
\left(\delta_{ij}\delta_{kl}+\delta_{ik}\delta_{jl}+\delta_{il}\delta_{jk}\right)\\
+\frac{1}{7}\Big(
\delta_{ij}\langle {n}_k{n}_l\rangle+\delta_{ik}\langle {n}_j{n}_l\rangle+\delta_{il}\langle {n}_j{n}_k\rangle\\
+\langle {n}_i{n}_j\rangle\delta_{kl}+\langle {n}_i{n}_k\rangle\delta_{jl}+\langle {n}_i{n}_l\rangle\delta_{jk}
\Big).
\label{eq13}
\end{multline}
The same closure, when applied to the isotropic outer distribution function $\Psi^{\mathrm{outer}}$ [Eq. (\ref{eq666})], reduces to:
\begin{multline}
\langle\boldsymbol{nnnn}\rangle^{\mathrm{outer}}=
\phi\oint 
(4\pi)^{-1} {n_in_jn_kn_l} d^2\boldsymbol{n}\\
=\frac{\phi}{15}\left(\delta_{ij}\delta_{kl}+\delta_{ik}\delta_{jl}+\delta_{il}\delta_{jk}\right).
 \label{eq14}
 \end{multline}
By combining Eqs.~(\ref{eq12}, \ref{eq14}), we finally arrive at the closed Gillissen-Wilson equation for microstructural evolution~\cite{gillissen2018modeling}:
\begin{multline}
\partial_t\langle \boldsymbol{nn}\rangle =
{\boldsymbol{L}}\cdot\langle \boldsymbol{nn}\rangle+\langle \boldsymbol{nn}\rangle\cdot{\boldsymbol{L}}^T-
2{\boldsymbol{L}}:\langle \boldsymbol{nnnn}\rangle\\
-\beta\left[
{\boldsymbol{E}}_e: \langle \boldsymbol{nnnn}\rangle+
\frac{
\color{black}
\phi
\color{black}
}{15}
 \left(2{\boldsymbol{E}}_c+\mathrm{Tr}({\boldsymbol{E}}_c)
 \boldsymbol{\delta}\right)
 \right],
  \label{eq15}
\end{multline}
in which $\langle \boldsymbol{nnnn}\rangle$ is now shorthand for the right hand side of Eq.~\eqref{eq13}.

A significant novelty of Eq.~(\ref{eq15}) 
is its separate linearity in the compressive and extensional components of the rate of strain tensor, 
making it overall non-linear in the rate of strain and thus distinct from various previous models 
that failed to adequately predict reversal flows without excessive parameters~\cite{chacko2018shear}. On reversal, 
the compressive and extensional components interchange so that contacts that were being pushed together are now pulled apart.

\subsection{Particle Stress and Contacts}
\label{stress}
A two-body approximation for the particle stress reads:
\begin{equation}
\boldsymbol{\Sigma}=n\langle \boldsymbol{Fr}\rangle,
\label{eq16}
\end{equation} 
where $\langle \cdots\rangle$ is the coarse-graining operator defined in Eq.~(\ref{eq7}), and $\boldsymbol{F}$ is the interparticle force. 
In the absence of tangential contact forces, this $\boldsymbol{F}$ equates to the last two terms of Eq.~(\ref{eq1}):
\begin{equation}
\boldsymbol{F}=C_2a\eta_s\left(\dot{\boldsymbol{r}}\cdot \boldsymbol{n}\right)\boldsymbol{n}-C_3a^2\eta_s\dot{\gamma}\Theta\left(2a-r\right)\boldsymbol{n}.
\label{eq17}    
\end{equation}
We continue to use this equation even in the presence of friction.
This might appear to be a drastic additional assumption but in fact our own data (generated with the DEM simulation introduced below), and also that of Ref. \cite{seto2018normal}, show that in the shear thickening range of volume fractions, tangential contributions to the stress remain subdominant. 
This subdominance does not contradict the fact that friction, by constraining tangential particle motion, greatly enhances normal contact forces. 
This enhancement is captured by Eq.~\eqref{eq1c} for $C_3$, which diverges at a jamming point that depends on both time-dependent microstructure (via $\xi$) and stress-dependent friction (via $\xi^J$).

Combining Eqs.~(\ref{eq4},\ref{eq7},\ref{eq16},\ref{eq17}) gives:
\begin{equation}
\boldsymbol{\Sigma}=
\eta_s \alpha \boldsymbol{E}:\langle\boldsymbol{nnnn}\rangle\\
-
\eta_s\chi \dot{\gamma}
\langle\boldsymbol{nn}\rangle_c,
\label{eq18}
\end{equation}
where $\langle\boldsymbol{nn}\rangle_c$ is the second order orientation moment 
of the contact part of $\Psi(\boldsymbol{r})$: 
\begin{equation}
\langle\boldsymbol{nn}\rangle_c=
\int_{r=2a}^{r=2a+\epsilon}\Psi\left(\boldsymbol{r}\right) \Theta(2a-r) \boldsymbol{nn}\, d^3\boldsymbol{r}.
\label{eq18e}
\end{equation}
The leading order behaviours of the pre-factors $\alpha$
and $\chi$ in Eq.~(\ref{eq18}) are found from Eqs.~(\ref{eq1a},\ref{eq1b},\ref{eq1c}) as:
\begin{equation}
\alpha(\phi)=\frac{\alpha_0}{\left(1-\phi/\phi^J_1\right)^{2}}\;\;, \;\;
\chi(\phi,\xi,\xi^J)=\frac{\chi_0}{\left(1-\xi/\xi^J\right)^{2}}. 
\label{eq18b}
\end{equation}
Here $\alpha_0$ and $\chi_0$ are fitting parameters, 
and $\xi^J$ depends on the particle pressure as specified in Eq.~(\ref{eq2}).

To obtain a closed form, the
contact moments $\langle\boldsymbol{nn}\rangle_c$
need to be approximated in terms of the coarse-grained moments $\langle\boldsymbol{nn}\rangle$ which includes all particle pairs in the coarse-graining shell.
To relate $\langle\boldsymbol{nn}\rangle_c$ to 
$\langle\boldsymbol{nn}\rangle$, we assume the following approximate parameterisation for $\Psi(\boldsymbol{r})$ within the coarse-graining shell: 
\begin{multline}
\Psi(\boldsymbol{r})=\Psi_{\boldsymbol{n}}(\boldsymbol{n})\Psi_{r}(r)\\
=\Psi_{\boldsymbol{n}}(\boldsymbol{n})a^{-2}\left[\epsilon^{-1}-2 C_4\frac{\boldsymbol{E}_c:\boldsymbol{nn}}{\vert\boldsymbol{E}_c\vert}\delta(r-2a)\right],
\label{eq18d}
\end{multline}
where 
$\Psi_{\boldsymbol{n}}(\boldsymbol{n})$ is the orientation distribution function 
(with no dependence on the radial distance),
$\Psi_{r}(r)$ is the radial distribution function (with no dependence on orientation),
$\delta(\cdots)$ is the Dirac delta function, $C_4$ is a pre-factor of order unity and 
$\vert\boldsymbol{E}_c\vert\equiv \sqrt{\boldsymbol{E}_c:\boldsymbol{E}_c}$.
Since closed contacts ($r=2a$) are predominantly oriented in a direction set by
$\boldsymbol{E}_c$, Eq.~(\ref{eq18d}) approximates 
the probability for closed contacts [$\Psi(r=2a)$]
with the probability for open contacts [$\Psi(r>2a)$],
weighted with the alignment of $\boldsymbol{n}$ in the 
compressive direction $-{\boldsymbol{E}_c:\boldsymbol{nn}}/{\vert\boldsymbol{E}_c\vert}$.
Combining Eqs.~(\ref{eq7},\ref{eq18d}) we find for the coarse-grained moments:
\begin{multline}
\langle\boldsymbol{nn}\rangle=\oint\Psi_{\boldsymbol{n}}(\boldsymbol{n})\boldsymbol{nn}\,d^2\boldsymbol{n}\\
-C_4\frac{\boldsymbol{E}_c}{\vert\boldsymbol{E}_c\vert}:\oint\Psi_{\boldsymbol{n}}(\boldsymbol{n})\boldsymbol{nnnn}\,d^2\boldsymbol{n}.
\label{eq18f}
\end{multline}
Assuming $\langle\boldsymbol{nn}\rangle_c\ll \langle\boldsymbol{nn}\rangle$ (see Fig.~\ref{figure4}b), 
we ignore the 
second term on the r.h.s. of Eq.~(\ref{eq18f}), and get:
\begin{equation}
\langle\boldsymbol{nn}\rangle=\oint\Psi_{\boldsymbol{n}}(\boldsymbol{n})\boldsymbol{nn}\,d^2\boldsymbol{n}.
\label{eq18g}
\end{equation}
Combining Eqs.~(\ref{eq18e},\ref{eq18d}) gives for the contact moments:
\begin{equation}
\langle\boldsymbol{nn}\rangle_c=
-C_4\frac{\boldsymbol{E}_c}{\vert\boldsymbol{E}_c\vert}:\oint\Psi_{\boldsymbol{n}}(\boldsymbol{n})\boldsymbol{nnnn}\,d^2\boldsymbol{n}.
\label{eq18h}
\end{equation}
By combining Eqs.~(\ref{eq18g},\ref{eq18h}),
we arrive at the following relation between 
the contact microstructure $\langle\boldsymbol{nn}\rangle_c$ 
and the coarse-grained microstructure $\langle\boldsymbol{nn}\rangle$:
\begin{equation}
\langle\boldsymbol{nn}\rangle_c=
-\frac{\boldsymbol{E}_c}{\vert\boldsymbol{E}_c\vert}:\langle\boldsymbol{nnnn}\rangle.
\label{eq20}
\end{equation}
Here we have set the proportionality constant $C_4$ to unity;
Eq.~(\ref{eq20}) thus identifies an approximated, non-normalised, contact microstructure 
that is calculable within our coarse-grained constitutive model.
Inserting Eqs.~(\ref{eq18b},\ref{eq20}) into Eq.~(\ref{eq18}) gives: 
\begin{equation}
\boldsymbol{\Sigma}=
\eta_s
\left[
{\frac{\alpha_0\boldsymbol{E}}{\left(1-\phi/\phi^J_1\right)^{2}} }
+{\frac{\chi_0 \boldsymbol{E}_c}{\left(1-\xi/\xi^J\right)^{2}}}
\right]
:\langle\boldsymbol{nnnn}\rangle,
\label{eq19}
\end{equation}
where the `jamming coordinate' 
\begin{equation}
\xi = \mathrm{Tr}\langle\boldsymbol{nn}\rangle_c,
\label{eq100}
\end{equation}
serves as 
a proxy for the coordination number $Z$ for direct particle contacts.
Combining Eqs. (\ref{eq20}, \ref{eq100}) gives:
\begin{equation}
\xi=-\frac{\langle\boldsymbol{nn}\rangle:\boldsymbol{E}_c}{\vert\boldsymbol{E}_c\vert}.
\label{eq21}
\end{equation}

Without a relation such as Eq.~\eqref{eq20}, the distance from jamming
is not deducible from the coarse-grained microstructure tensor 
$\langle\boldsymbol{nn}\rangle$: 
a proxy of some sort is essential for our constitutive model of shear thickening to be closed at coarse-grained level. 
However, Eq.~(\ref{eq21}) comprises a relatively crude approximation; 
some other combination of $\langle\boldsymbol{nn}\rangle$ and flow tensors might approximate $Z$ more accurately. Indeed it is found in particle-based simulations that the reduced viscosity $\eta_r$ in steady shear flow 
has a different power-law dependence on each: $(1-\xi/\xi^J)^{-2}\sim\eta_r\sim(1-Z/Z^J)^{-4}$~\cite{gillissen2019constitutive}.  
Moreover we will see in Fig.~\ref{figure6} below that $\langle\boldsymbol{nn}\rangle_c$ 
has some shortcomings when compared with the results of particle-based simulations.

With this in mind, although $\xi$ was constructed above as an estimator of $Z$, we note that its conceptual role in our constitutive model does not require this interpretation. Instead it can be viewed as a microstructural scalar that can capture the distance from a jamming point, $\xi-\xi^J$, in time-dependent flows, just as $\phi-\phi^J$ does in the Wyart-Cates theory for steady flow~\cite{gillissen2019constitutive}.
The jamming coordinate $\xi$ thereby emerges as the central variable to model shear thickening:
in Eq.~(\ref{eq2}) the stress is assumed to diverge when $\xi$ 
reaches a critical value $\xi^J$, that smoothly reduces from a 
larger frictionless value $\xi^J_1$, to a smaller frictional value $\xi^J_2$, 
when the pressure $\Pi$ in the system exceeds the onset value $\Pi^*$.

\subsection{Determination of Parameters}
\label{parameters}
The critical values $\xi^J_1$ and $\xi_2^J$ are 
found by demanding that in steady shear 
frictionless and frictional jamming occur at volume fractions 
$\phi^J_1$ and 
$\phi^J_2$, respectively.
For steady $xy$ shear flow the solution 
to Eq.~(\ref{eq15}) reads:
\begin{multline}
\langle\boldsymbol{nn}\rangle=\frac{\phi}{\left(9 \beta ^2+54 \beta +416\right)}\times \\
\Big\{
\tfrac{1}{15}{\left(129 \beta ^2-374 \beta +3256\right)}\boldsymbol{\delta}_1\boldsymbol{\delta}_1\\ 
-\tfrac{28}{5}{\left(\beta ^2 -3 \beta   \right)}\left(\boldsymbol{\delta}_1\boldsymbol{\delta}_2+\boldsymbol{\delta}_2\boldsymbol{\delta}_1\right)\\ 
+\tfrac{1}{15}{\left(129   \beta ^2+410\beta +904 \right) }\boldsymbol{\delta}_2\boldsymbol{\delta}_2\\
+\tfrac{1}{15}{\left(87   \beta ^2+564   \beta +820 \right) }\boldsymbol{\delta}_3\boldsymbol{\delta}_3 
\Big\},
\label{eq30}
\end{multline}
where we recall that $\beta$ depends on $\phi$ [Eq. (\ref{eq999})].
Inserting Eq.~(\ref{eq30}) into Eq.~(\ref{eq21}) gives for the jamming coordinate in steady shear:
\begin{equation}
\xi=\phi\frac{\left(213 \beta ^2-234 \beta +2080\right) }{15 \left(9 \beta ^2+54 \beta +416\right)}.    
\label{eq33}
\end{equation}
Eq. (\ref{eq33}) shows that in steady simple shear flow, 
$\xi$ is proportional to the volume fraction $\phi$ 
which follows from the assumption [Eq. (\ref{eq14})] that the outer distribution of the pair separation vector 
is proportional to $\phi$.

Requiring that frictionless and frictional jamming occur at volume fractions 
$\phi_1^J$ and $\phi^J_2$ demands the following critical values for 
the frictionless and frictional jamming coordinates:
\begin{equation}
\xi^J_{1,2}=\phi^J_{1,2}\frac{\left(213 \beta ^2-234 \beta +2080\right) }{15 \left(9 \beta ^2+54 \beta +416\right)}.
\label{eq34}
\end{equation}

Eqs.~(\ref{eq2},\ref{eq999},\ref{eq15},\ref{eq18b},\ref{eq19},\ref{eq21},\ref{eq34}) form a closed system 
for the microstructure and stress. For any given volume fraction $\phi$, the model contains
parameters $\alpha_0,\beta_0$, $\chi_0$, $\Pi^*$, $\phi^J_1$ and $\phi_2^J$. 
Of these parameters, $\phi^J_1$ and $\phi_2^J$ are directly determinable 
from experimental or computational data pertaining the dependence of the viscosity on the volume fraction under frictionless and frictional conditions, respectively,
and $\Pi^*$ enters only through the scale factor relating the reduced shear rate $\dot\gamma_r$ [Eq. (\ref{eq10})] to the actual one, $\dot\gamma$. In previous work we used steady state microstructural and viscosity data (for various $\phi$), and microstructural reversal data (for $\phi = 0.56$), from particle-based simulations, to fit 
$\phi_1^J = 0.65, \phi_2^J = 0.57$, $\Pi^* = 0.037F^*/a^2$, 
$\alpha = 120$, $\beta = 50$ and $\chi_0 = 2.4$,
which for $\phi=0.56$ correspond to $\alpha_0 = 2.3$ and $\beta_0 = 6.9$. 
It is noted that these $\phi_{1,2}^J$ differ slightly from the values 
extrapolated from the DEM data $\phi_1^J = 0.644, \phi_2^J = 0.578$.
In Ref. \cite{gillissen2019constitutive}, the model was then 
used to predict, without further parameter fitting, the rheological reversal data at $\phi = 0.56$, with qualitatively good agreement in most respects~\cite{gillissen2019constitutive}.  
It is also noted that according to Eq. (\ref{eq34}) $\xi^J_{1,2}\approx 0.88, 0.78$, 
while the corresponding coordination numbers are $Z^J_{1,2}=6,4$. 
It is therefore re-emphasised that, although $\xi$ might be interpreted as an approximation for $Z$, 
these parameters differ numerically, and they are not linearly proportional. 
They nevertheless play similar roles, in providing the distance to the jamming point.

We next briefly review the particle-based simulation methodology before making a similar comparison of the constitutive model with a contrasting type of flow in which transverse oscillations are superposed onto steady shearing. 

\section{Discrete-Element Model}
\label{particle}
Our discrete-element method (DEM) simulation model considers non-Brownian, almost non-inertial, neutrally buoyant particles in a periodic cubic box at volume fraction $\phi$. The particles are an equimolar mixture of radii $a$ and $1.4a$, and have density $\rho$. The box is initialised with 1500 nonoverlapping particles and we report averages over 10 realisations.
The simulation box (volume $V$) is deformed with a superposition of a steady shear flow (rate $\dot{\gamma}$) 
and a transverse oscillating shear flow (amplitude $\gamma$ and frequency $\omega$) 
with a velocity gradient $\boldsymbol{L}$ and rate of strain tensor $\boldsymbol{E}$ that
are given by Eqs.~(\ref{eq22},\ref{eq25}) below, respectively.
The nondimensional control parameters for this family of flows are, when applied to shear-thickening suspensions, 
the volume fraction
$\phi$, the oscillation strain amplitude $\gamma$, 
the dimensionless oscillation frequency $\dot{\gamma}_\perp$ [Eq. (\ref{eq40})]
and the dimensionless shear rate $\dot{\gamma}_r$ [Eq. (\ref{eq10})].
For the transverse flow to be effective at reducing the viscosity,
its amplitude must be large enough to break direct contacts
yet small enough to inhibit significant contact formation in $yz$.
Within this range (approximately $10^{-4}<\gamma<0.05$) the results are almost independent of $\gamma$~\cite{lin2016tunable}, and in the following we fix $\gamma=0.01$.

Hydrodynamic interactions between particles are computed as described in Refs.~\cite{jeffrey1984calculation,jeffrey1992calculation,kim1991microhydrodynamics,ball1997simulation}.
For neighbouring particles 1 and 2, translating with velocities $\boldsymbol{U}_1$, $\boldsymbol{U}_2$ and rotating at $\boldsymbol{\Omega}_1$, $\boldsymbol{\Omega}_2$, and with centre-centre vector $\boldsymbol{r}$ (and $\boldsymbol{n}=\boldsymbol{r}/|\boldsymbol{r}|$) pointing from particle 2 to particle 1, the force $\boldsymbol{F}^h$ and torque $\boldsymbol{\Gamma}^h$ on particle 1 are given by:
\begin{subequations}
\begin{equation}
\begin{split}
{\bm F}^h/\eta_s =& 
\left[X^A_{11} \boldsymbol{nn} + Y_{11}^A({\bm \delta}-\boldsymbol{nn})\right]\cdot({\bm U}_2 - {\bm U}_1)\\
&+ Y^B_{11}({\bm \Omega}_1 \times {\bm n})+Y^B_{21}({\bm \Omega}_2 \times {\bm n}) \text{,}
\end{split}
\end{equation}
\begin{equation}
\begin{split}
{\bm \Gamma}^h/\eta_s =& Y^B_{11}({\bm U}_2 - {\bm U}_1)\times {\bm n}\\
&- ({\bm \delta}-\boldsymbol{nn})\cdot(Y_{11}^C{\bm \Omega}_1 + Y_{12}^C{\bm \Omega}_2) \text{,}
\end{split}
\end{equation}
\label{lubric}
\end{subequations}
where $\eta_s$ is the solvent viscosity.
The surface-surface separation is given, for particle radii $a_1$ and $a_2$, by $h = |{\bm r}| - (a_1 + a_2)$, which is nondimensionalised as $2h/(a_1+a_2)$. The scalar resistances $X^A_{11}$, $Y^A_{11}$, $Y^B_{11}$, $Y^B_{21}$, $Y^C_{11}$ and $Y^C_{12}$ are given elsewhere~\cite{cheal2018rheology}. We neglect interactions that have $h>0.05a$.
A drag force and torque act on particle 1 at position ${\bm x}_1$, given by 
\begin{subequations}
\begin{equation}
{\bm F}^{d} = -6\pi \eta_sa_1({\bm U}_1 - {\bm U}({\bm x}_1)) \text{,}
\end{equation}
\begin{equation}
{\bm \Gamma}^{d} = -8\pi \eta_s a_1^3({\bm \Omega}_1 - {\bm \Omega}({\bm x}_1)) \text{,}
\end{equation}
\end{subequations}
with ${\bm \Omega}=\tfrac{1}{2}\boldsymbol{\nabla}\times\boldsymbol{U}$ the fluid vorticity vector,
and the streaming velocity given by ${\bm U}({\bm x}) = {\bm L}\cdot{\bm x}$.

Below a separation $h_\text{min}=0.001a$, hydrodynamic forces are regularised and particles enter into direct contact. Particle pairs with overlap $\delta = ((a_1+a_2)-|{\bm r}|)\Theta((a_1+a_2)-|{\bm r}|)$ (with Heaviside function $\Theta$) and centre-centre unit vector ${\bm n}$ lead to contact force and torque on particle 1 according to \cite{cundall1979discrete}:
\begin{subequations}
\begin{equation}
{\bm F}^c = k_n\delta{\bm n} - k_t{\bm t} \text{,}
\end{equation}    
\begin{equation}
{\bm \Gamma}^c = a_1 k_t({\bm n} \times {\bm t}) \text{,}
\end{equation}
\label{cundall}
\end{subequations}
where ${\bm t}$ represents the incremental tangential displacement, reset at the initiation of each contact. Here $k_n$ and $k_t$ are stiffnesses, with $k_t = (2/7)k_n$. The tangential force component is restricted by a friction coefficient $\mu=1$ so that $|k_t{\bm t}| \leq \mu k_n\delta$.
Stress-dependence enters through $\mu$, following Ref.~\cite{mari2014shear}:
\begin{equation}
    |k_t{\bm t}| \leq  \left\{
                \begin{array}{ll}
                  \mu k_n(\delta-\delta^*) & \text{for } \delta>\delta^*\\
                  0 & \text{otherwise}
                \end{array}
              \right.
\end{equation}
where $F^* \equiv k_n\delta^*$ is the normal force above which friction is activated, 
leading to a nondimensional shear rate 
$\dot{\gamma}_r=\dot{\gamma}\eta_s/\Pi^*\sim \dot{\gamma}\eta_sa^2/F^*$. 

Particle trajectories are computed from the above forces, and the components of the stress tensor $\boldsymbol{\Sigma}$ are calculated by summing $-\boldsymbol{Fr}$ over all interacting particle pairs and dividing by $V$.
The contact microstructure is computed as $\langle\boldsymbol{nn}\rangle_c$, where 
$\langle{\cdots}\rangle_c$ denotes averaging over all particle pairs for which the contact forces [Eq.(~\ref{cundall})] are activated. 
We also construct a coarse-grained microstructure 
$\langle\boldsymbol{nn}\rangle$,
where $\langle{\cdots}\rangle$ averages over all particle pairs
that interact via direct contact forces or lubrication forces, the 
latter being cut off beyond a separation distance of $h=0.05 a$. 
Below we will compare these quantities to constitutive model predictions. 
In addition to the control parameters described above, the model leads to a Stokes number $\text{St}=\rho\dot{\gamma}a^2/\eta_s$ and a $k_n$-scaled shear rate $\hat{\dot{\gamma}} = 2\dot{\gamma}a/\sqrt{k_n/(2 \rho a)}$.
We set $\mathrm{St}<10^{-3}$ 
and 
$\hat{\dot{\gamma}} <10^{-5}$
to approximate inertia-free, hard sphere conditions.
The model is implemented in $\texttt{LAMMPS}$~\cite{plimpton1995fast}.

\section{Results}
\label{results}

We now test the microstructure and stress predicted 
by our constitutive model against 
data generated by the DEM
simulation
at volume fraction $\phi=0.56$,
under a homogeneous, time-dependent velocity gradient
\begin{equation}
\boldsymbol{L}=
\left(
\begin{array}{ccc}
0 & \dot{\gamma}&0 \\
0 & 0 & 0 \\
0 & \omega\gamma\cos(\omega t)&0\\
\end{array}\text{,}
\right),
\label{eq22}
\end{equation}
corresponding to a deformation rate
\begin{equation}
{\bm E} = \begin{pmatrix}
    0 & \frac{1}{2}\dot{\gamma} & 0  \\
    \frac{1}{2}\dot{\gamma} & 0& \frac{1}{2}\gamma\omega\cos{(\omega t)}  \\
    0 & \frac{1}{2}\gamma\omega\cos{(\omega t)}                    & 0 
  \end{pmatrix}\text{.}
\label{eq25}
\end{equation}
In the limit of large $\dot{\gamma}_{\perp}$, 
we have that: 
\begin{equation}
\boldsymbol{E}_e=\tfrac{1}{2}\gamma\omega\vert\cos(\omega t)\vert\boldsymbol{n}_e\boldsymbol{n}_e,\hspace{1cm}
\boldsymbol{E}_c=-\tfrac{1}{2}\gamma\omega\vert\cos(\omega t)\vert\boldsymbol{n}_c\boldsymbol{n}_c,
\label{eq41}
\end{equation}
where $\pm\tfrac{1}{2}\gamma\omega\vert\cos(\omega t)\vert$ are the expansive and compressive eigenvalues of $\boldsymbol{E}$ of Eq.(\ref{eq25}) 
and $\boldsymbol{n}_e=x(0,1,1)/\sqrt{2}+(1-x)(0,-1,1)/\sqrt{2}$ and 
$\boldsymbol{n}_c = (1-x)(0,1,1)/\sqrt{2} + x(0,-1,1)/\sqrt{2}$ are the corresponding eigenvectors,
with $x=\Theta[\cos(\omega t)]$ and $\Theta(\cdot)$ the Heaviside step function.
Note that these eigenvectors interchange direction after each half oscillation period 
and, on average, $-\overline{E}_c=\overline{E}_e\sim \gamma\omega\left(\boldsymbol{\delta}_y\boldsymbol{\delta}_y+\boldsymbol{\delta}_z\boldsymbol{\delta}_z\right)$.

Setting $\gamma=0.01$ and $\phi=0.56$, the remaining control parameters 
are $\dot{\gamma}_r$ [Eq. (\ref{eq10})] and $\dot{\gamma}_\perp$ [Eq. (\ref{eq40})] which quantify the influence, respectively,
of frictional contact forces and transverse oscillations.
Below we first focus on the limiting cases of $\dot{\gamma}_\perp=0$ and $\dot{\gamma}_\perp=\infty$,
before considering the behaviour of shear-thickened suspensions ($\dot{\gamma}_r\gg1$) at intermediate values of $\dot{\gamma}_\perp$.
We finally present full maps of the viscosity as functions
of $\dot{\gamma}_r$ and $\dot{\gamma}_\perp$.

\subsection{Steady behaviour with $\dot{\gamma}_{\perp}=0$}
\begin{figure}
\centerline{
\includegraphics[trim = 0mm 0mm 0mm 0mm,clip,width=0.45\linewidth,page=1]{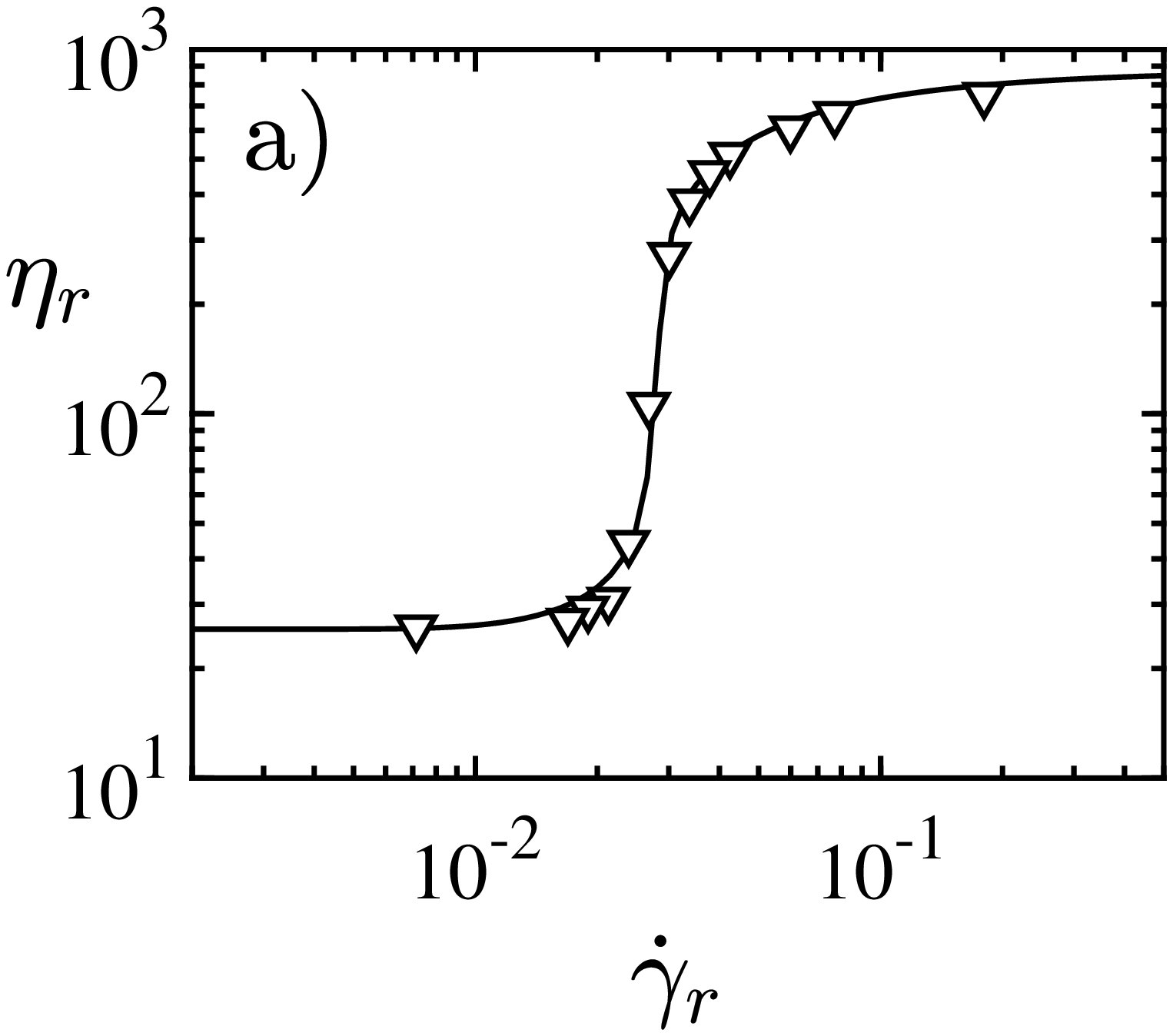}
\includegraphics[trim = 0mm 0mm 0mm 0mm,clip,width=0.54\linewidth,page=1]{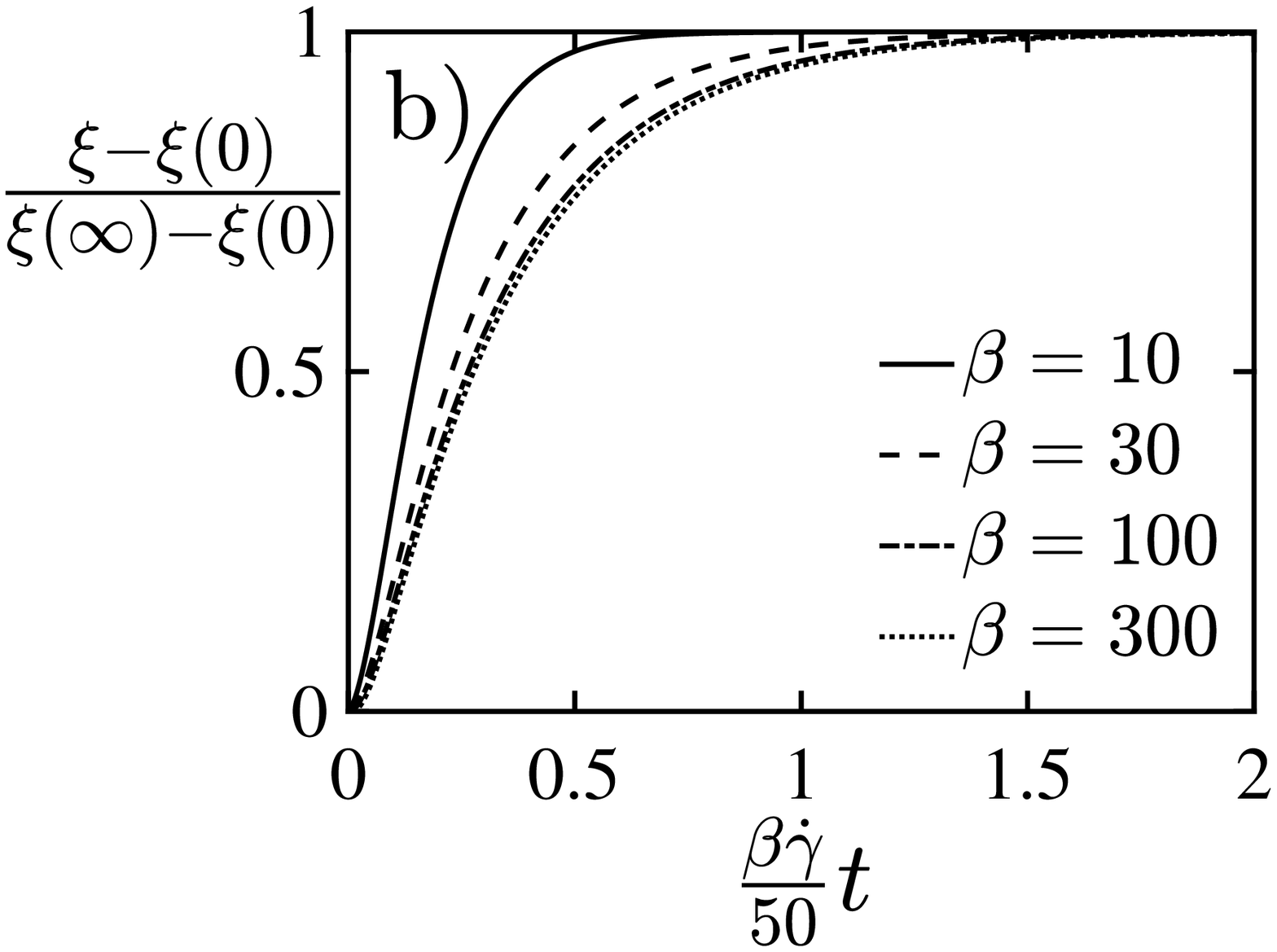}
}
\caption{
(a)
Relative suspension viscosity $\eta_r$ 
in steady shear as a function of dimensionless shear rate $\dot{\gamma}_r$,
measured in the absence of transverse oscillations, $\dot{\gamma}_\perp=0$. 
Shown are results from the constitutive model (solid line) and discrete-element simulation (triangles) at $\phi=0.56$.
(b) Recovery of the jamming coordinate $\xi$ after shear reversal 
scaled with its values at $t=0$ and at $t=\infty$,
for various values of the microstructure association rate $\beta$.
}
\label{figure2}
\end{figure}

A flow curve for steady shear without transverse oscillation ($\dot{\gamma}_\perp=0$) is shown in
Fig.~\ref{figure2}a, demonstrating good agreement in the viscosity prediction of the DEM simulation and the constitutive model. 
The parameter values are those chosen in~\cite{gillissen2019constitutive} as detailed in Sec.~\ref{parameters} above.
The constitutive model predicts
for the shear component of the coarse-grained microstructure that
$\nn_{xy}<0$ and
for the normal components that $\nn_{yy}>\nn_{xx}>\nn_{zz}$, 
and similar behaviour for the contact microstructure $\langle\boldsymbol{nn}\rangle_{c,ij}$. 
The model thus predicts 
a positive first normal stress difference
$\zeta_1=(\Sigma_{xx}-\Sigma_{yy})/\Sigma_{xy}$ and 
a negative second normal stress difference
$\zeta_2=(\Sigma_{yy}-\Sigma_{zz})/\Sigma_{xy}$.
This is in partial agreement with DEM, which predicts 
that $\nn_{xy}<0$ and that $\nn_{xx}>\nn_{yy}>\nn_{zz}$, 
and similar behaviour for $\langle\boldsymbol{nn}\rangle_{c,ij}$. 
Correspondingly DEM predicts that $\zeta_1<0$ (but very small) and 
$\zeta_2<0$.
In general the constitutive model overestimates the microstructural anisotropy
and $\vert\zeta_{1,2}\vert$, as compared to the DEM simulation~\cite{gillissen2019constitutive}.
Further results and discussion relating to the steady shear stress and the microstructure
predicted by our model are given in Ref.~\cite{gillissen2019constitutive}.

\subsection{Shear reversal with $\dot{\gamma}_{\perp}=0$}
Figure \ref{figure2}b shows the recovery of the jamming coordinate $\xi$ [Eq. (\ref{eq21})] after shear reversal for various values of the microstructure association rate $\beta$. 
In this case, the suspension is subjected to a negative $xy$-shear flow, without $zy$-shear oscillations.
When the steady state is reached, the shear flow is reversed at $t=0$.
It is seen that for $\beta\gtrsim 30$ the recovery shear rate $\approx \beta\dot{\gamma}/50$ and 
full recovery is achieved after a strain of $\approx 50/\beta$.
This suggests that 
in the constitutive model
the (transverse) oscillatory strain $\gamma=10^{-2}$ is unable to induce 
significant microstructural reorganisation, 
for the present case, where $\beta=50$.

\subsection{Limiting behaviour for large $\dot{\gamma}_\perp$}
\label{limit}
Next, we consider $\langle\boldsymbol{nn}\rangle$ predicted by the constitutive model, 
in the limit of a very large oscillation frequency
$\dot{\gamma}_{\perp}\gg 1$.
In this limit, $\langle\boldsymbol{nn}\rangle$ is dominated by the oscillatory flow, 
while the steady shear component only presents an 
$\mathcal{O}(\dot{\gamma}_{\perp}^{-1})$ 
perturbation to $\langle\boldsymbol{nn}\rangle$.  
We have seen in Fig. \ref{figure2}b, that substantial microstructural reorganisation requires 
$\beta\gamma/50\gtrsim 1$, which is not met by current conditions, in which
$\beta=50$ and $\gamma=10^{-2}$.
Under present conditions, 
the periodic changes in $\langle\boldsymbol{nn}\rangle$ are 
$\mathcal{O}(\beta\gamma/50)$,  
while on average $\langle\boldsymbol{nn}\rangle$ remains isotropic and 
equilibrated to the external microstructure 
$\langle\boldsymbol{nn}\rangle=\langle\boldsymbol{nn}\rangle^{\mathrm{outer}}+\mathcal{O}(\beta\gamma/50)$, where 
$\langle\boldsymbol{nn}\rangle^{\mathrm{outer}}=
\phi\boldsymbol{\delta}/3$.

The nearly isotropic $\langle\boldsymbol{nn}\rangle$ at $\dot{\gamma}_{\perp}\gg 1$
corresponds to a jamming coordinate of $\xi\approx \phi/3$ [found by inserting $\langle\boldsymbol{nn}\rangle=\phi\boldsymbol{\delta}/3$ in Eq.~(\ref{eq21})], 
roughly four times smaller than $\xi\approx 1.4\phi$, 
which follows from 
inserting the steady shear microstructure [Eq. (\ref{eq30})] into 
Eq. (\ref{eq21}) and using our chosen model parameter $\beta=50$.

If one now imposes a small steady shear flow perpendicular to this oscillatory state, 
the ability to flow in the steady direction is governed by the time-averaged $\langle\boldsymbol{nn}\rangle$ 
which is isotropised by the dominant oscillatory flow. 
This isotropisation 
corresponds to a lower $\xi\approx \phi/3$ as compared to that in steady shear $\xi\approx 1.4\phi$, 
taking the system further from jamming, 
thereby causing a reduction in the modelled stress, via Eq.~(\ref{eq19}); 
for a system close to the steady-shear jamming point, this reduction can be arbitrarily large. This explanation of the unjamming effect of transverse oscillation in the large $\dot\gamma_\perp$ limit is broadly consistent with previous discussions~\cite{lin2016tunable,ness2018shaken}.

In what follows, we solve the full constitutive model numerically, across a wide range of $\dot\gamma_\perp$. 
We thereby confirm that 
for very large $\dot\gamma_\perp$, 
the model predicts an isotropic coarse-grained microstructure 
${\langle \boldsymbol{nn}\rangle}\approx\phi\boldsymbol{\delta}/3$ [Fig.~\ref{figure4}c below]
with $\xi \approx \phi/3$. 

\subsection{Role of $\dot{\gamma}_\perp$: transient behaviour}

\begin{figure}[b]
\centerline{
\hspace{0.3cm}
\psfig{file=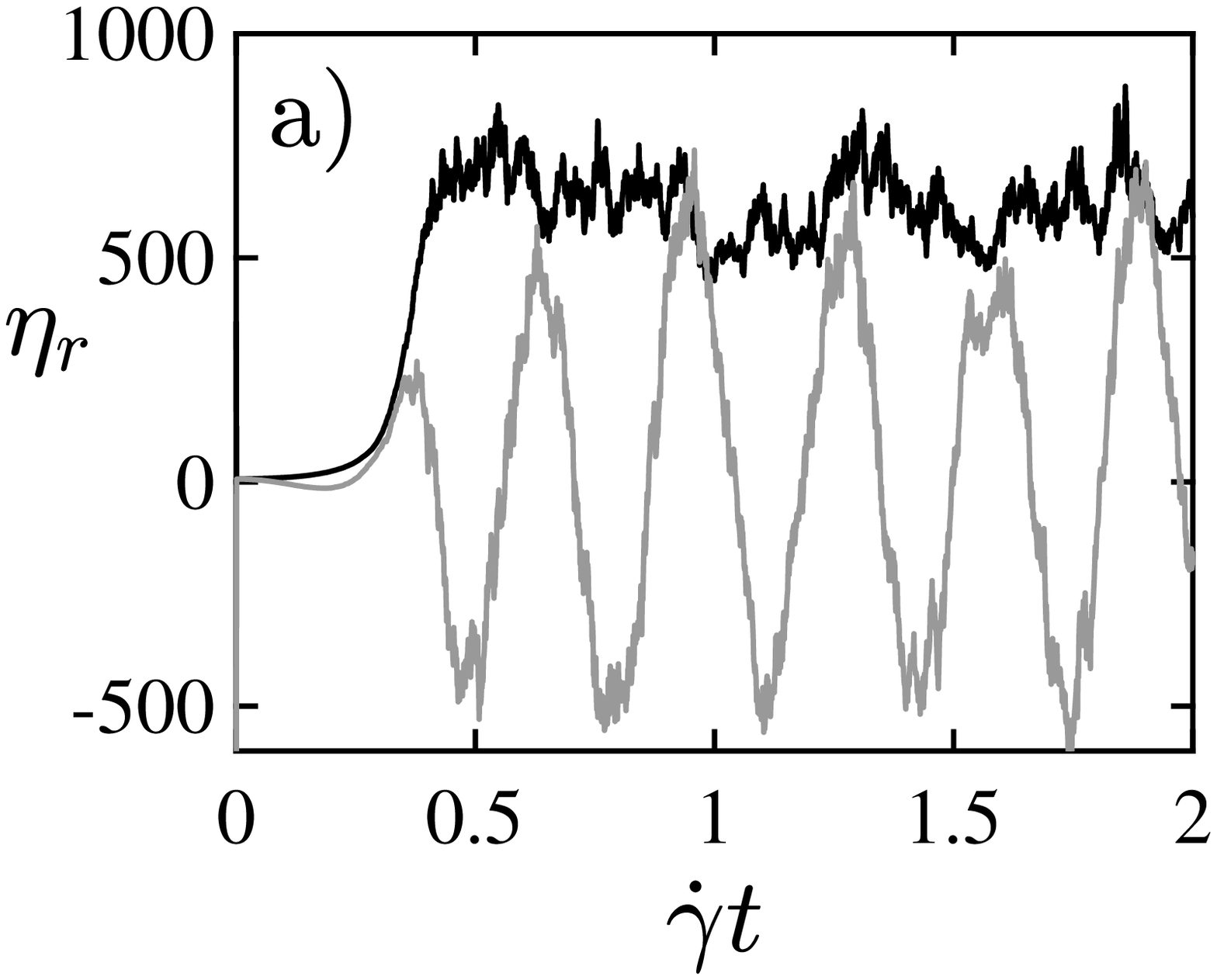,width=0.47\linewidth}\hspace{0.5cm}
\psfig{file=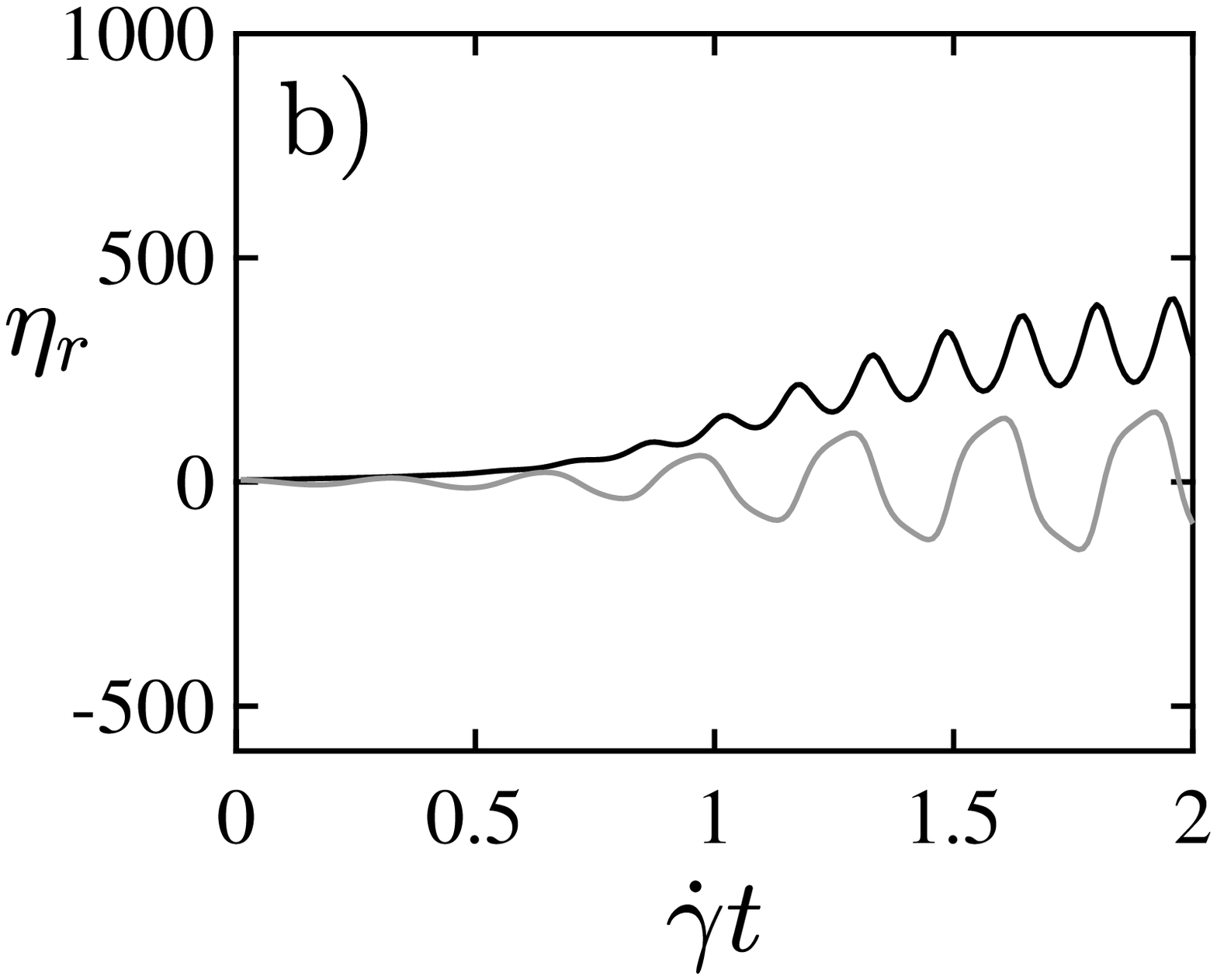,width=0.47\linewidth}
}
\centerline{
\psfig{file=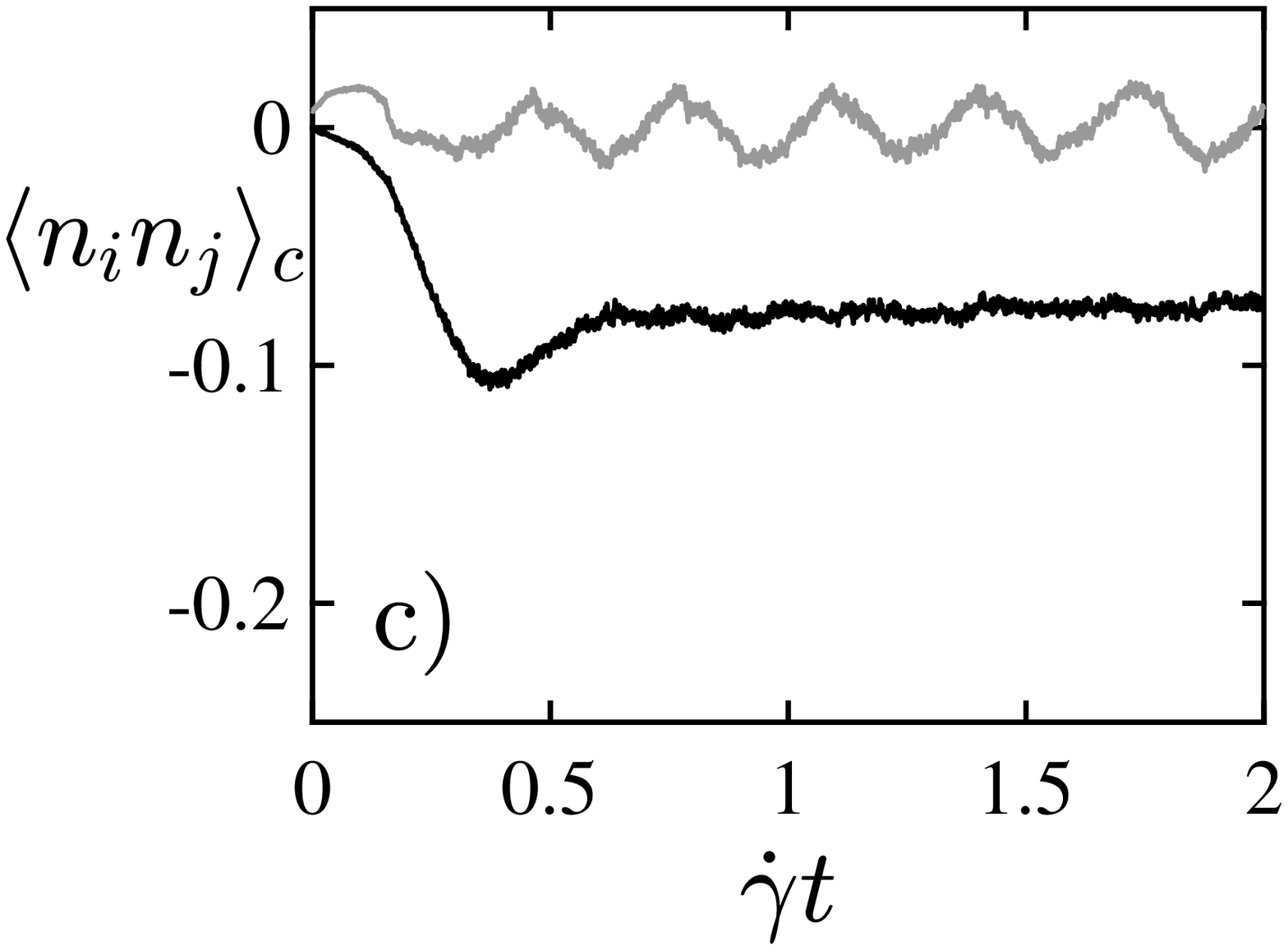,width=0.5\linewidth}
\psfig{file=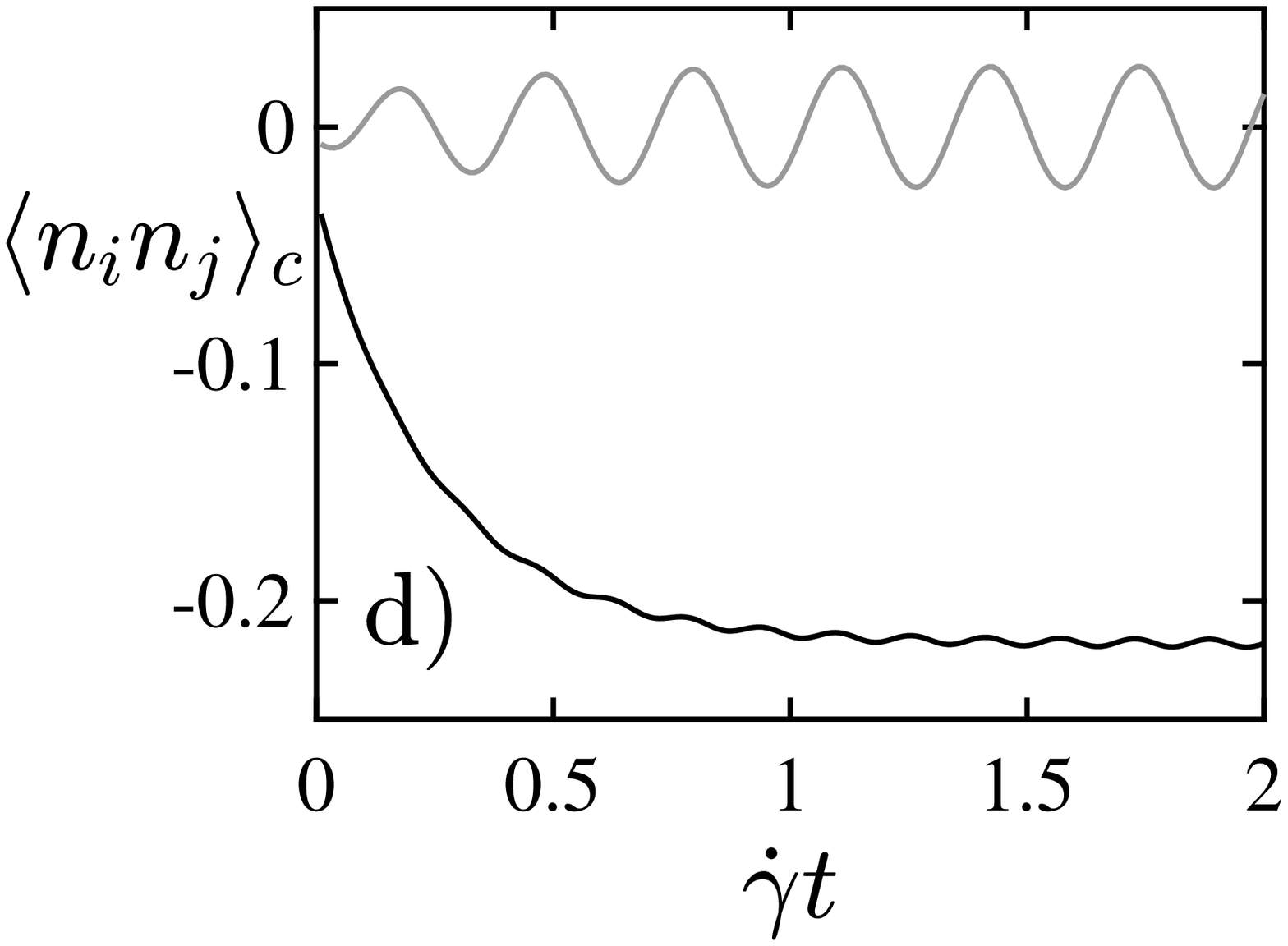,width=0.5\linewidth}
}
\caption{
Transient response of DEM simulation 
and constitutive model 
to shear flow with superposed transverse oscillations 
for $\dot{\gamma}_r=\infty$ and $\dot{\gamma}_{\perp}=0.2$.
Shown are steady shear viscosity $\Sigma_{xy}/(\dot{\gamma}\eta_s)$ (black)
and oscillatory viscosity component $\Sigma_{yz}/(\omega{\gamma}\eta_s)$ (grey) 
for DEM (a) and constitutive model (b),
and contact microstructure components $\langle\boldsymbol{nn}\rangle_{c,xy}$ (black) and 
$\langle\boldsymbol{nn}\rangle_{c,yz}$ (grey)
for DEM (c) and constitutive model (d).
}
\label{figure3}
\end{figure}

\begin{figure}[b]
\centerline{
\psfig{file=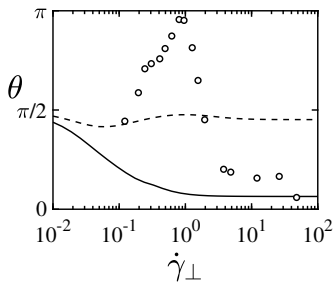,width=0.5\linewidth}
}
\caption{
Phase angle $\theta$
for the coarse-grained microstructure component 
$\langle \boldsymbol{nn}\rangle_{yz}$ in the constitutive model (solid line),
for the contact microstructure component $\langle \boldsymbol{nn}\rangle_{c,yz}$ in the constitutive model (dashed line) and
for the contact microstructure component $\langle \boldsymbol{nn}\rangle_{c,yz}$ in DEM simulation (markers). 
}
\label{figure6}
\end{figure}

We next compare results for intermediate values of $\dot{\gamma}_\perp$,
focussing again on the shear-thickened case, $\Pi\gg\Pi^*$. This case is described by the limit $\dot{\gamma}_r=\infty$ where frictional contacts are maximized [$\Pi^*=0$ and $f = 1$ in Eq. (\ref{eq2})] so that the role of particle-particle contact forces, at least under steady shear flow,  is maximally important.
We first present the behaviour observed in discrete-element simulations before discussing the constitutive model predictions.

Shown in Figs.~\ref{figure3}a, c are examples of time series
for the steady shear stress $\Sigma_{xy}$,
and the transverse one $\Sigma_{yz}$,
as well as the corresponding components 
of the contact microstructure tensor, $\langle\boldsymbol{nn}\rangle_{c,xy}$ and 
$\langle\boldsymbol{nn}\rangle_{c,yz}$, as
obtained by DEM simulations with
$\dot{\gamma}_r = \infty$ and
$\dot{\gamma}_\perp=0.2$.
Starting from a contact-free state, the steady shear flow component leads to a gradual building of particle contacts,
predominantly oriented along the compressive direction of the steady shear. 
This process results in a large shear stress $\Sigma_{xy}$ and a negative contact microstructure component $\langle\boldsymbol{nn}\rangle_{c,xy}$.
(Note that the definition of $\langle\boldsymbol{nn}\rangle_{c,xy}$ is such that it is negative under $xy$-shear flow with positive $\partial_y u_x$.)
Meanwhile the transverse shear generates oscillations in $\Sigma_{yz}$ in phase with the oscillatory shear rate $\gamma\omega\cos(\omega t)$, 
whose amplitude increases during the first few cycles as the steady flow component generates contacts.
The transverse component of the contact microstructure $\langle\boldsymbol{nn}\rangle_{c,yz}$ oscillates in anti-phase 
with the oscillatory shear rate,
which is understood by noting that at $\dot{\gamma}_\perp=0.2$
the microstructural response is sufficiently fast compared to the change in flow direction
that we essentially have a series of steady-state shear flows with a slowly changing direction~\cite{lin2016tunable,ness2018shaken}.

In Figs.~\ref{figure3}b, d we show time series of the stress components 
$\Sigma_{xy}$, $\Sigma_{yz}$ and of the 
contact microstructure components 
$\langle \boldsymbol{nn}\rangle_{c,xy}$ and $\langle \boldsymbol{nn}\rangle_{c,yz}$,
predicted by the constitutive model with 
$\dot{\gamma}_r = \infty$ and
$\dot{\gamma}_\perp=0.2$.
The constitutive model agrees qualitatively with the discrete-element model (Figs.~\ref{figure3}a, c). 
The contact microstructure $\langle \boldsymbol{nn}\rangle_{c,xy}$ develops over a few steady strain units, 
accompanied by substantial growth of $\Sigma_{xy}$ and $\Sigma_{yz}$. After the initial transient,
$\langle \boldsymbol{nn}\rangle_{c,xy}$ is nearly steady and negative,
while $\langle \boldsymbol{nn}\rangle_{c,yz}$ oscillates in anti-phase to the transverse shear rate $\gamma\omega\cos(\omega t)$. 
Quantitatively, the constitutive model requires a larger strain for $\Sigma_{xy}$ to develop fully, 
and it does not capture the transient peak in $\langle\boldsymbol{nn}\rangle_{c,xy}$.

However, while the transverse viscosity response found by DEM simulation remains
in phase with the transverse shear rate for all $\dot{\gamma}_\perp$,
the phase angle of $\langle\boldsymbol{nn}\rangle_{c,yz}$ 
shows a nonmonotonic dependence on $\dot{\gamma}_\perp$.
Shown in Fig.~\ref{figure6} are the phase angle $\theta$ 
of $\langle\boldsymbol{nn}\rangle_{c,yz}$ relative to minus the $zy$-strain ($-\gamma\sin(\omega t)$)
found by DEM simulation,  and the same phase angle for both $\langle \boldsymbol{nn}\rangle$ and $\langle \boldsymbol{nn}\rangle_c$ in the constitutive model.
The DEM simulations show that $\theta$ transitions as a function of $\dot{\gamma}_{\perp}$
from (i) $\theta\approx \pi/2$ (anti-phase with the $zy$-shear rate), 
via (ii) $\theta\approx \pi$ (in-phase with $zy$-strain),
to (iii) $\theta\approx 0$ (anti-phase with the $zy$-strain).
The physics of this sequence is explored in detail elsewhere~\cite{lin2016tunable}. 
Briefly, the three regimes correspond to (i) instant adaptation, where the contact microstructure tensor tracks the velocity gradient tensor as this oscillates around its mean value in a quasi-steady-state fashion; (ii) chain tilting, where the oscillatory flow deforms contacts faster than they are replaced by new ones but does not break up force chains; and (iii) chain breaking where the flow-induced contact network of the steady shear is substantially disrupted by the transverse oscillation. 

The constitutive model 
predicts different behaviours of the phase angle depending on whether the contact microstructure  $\langle \boldsymbol{nn}\rangle_c$ or the coarse-grained microstructure $\langle \boldsymbol{nn}\rangle$ is considered. The first of these shows $\theta\approx \pi/2$ over the entire $\dot{\gamma}_{\perp}$-range and is quite unlike the DEM data.
Interestingly, this discrepancy is inherent in the definition of 
$\langle\boldsymbol{nn}\rangle_c$ in Eq.~(\ref{eq20}). It follows from this
definition that the oscillations in $\langle\boldsymbol{nn}\rangle_c$ 
must remain almost in anti-phase with the oscillations in $\boldsymbol{E}$. 
This is readily seen in the limit $\dot{\gamma}_{\perp}=\infty$, 
where $\boldsymbol{E}_c= E_{c,yz}(\boldsymbol{\delta}_y\boldsymbol{\delta}_z+\boldsymbol{\delta}_z\boldsymbol{\delta}_y)$ 
since $E_{c,xy}/E_{c,yz}=\dot{\gamma}_{\perp}^{-1}=0$,
and 
$\langle\boldsymbol{nn}\rangle\approx\phi\boldsymbol{\delta}/3$
(see Sec. \ref{limit}). Inserting these expressions and Eq.~(\ref{eq13}) into Eq.~(\ref{eq20}) gives
$\langle\boldsymbol{nn}\rangle_{c,yz}=-(\phi/15)E_{c,yz}/\vert E_{c,yz}\vert$, which is in anti-phase 
with $E_{c,yz}$.
This phase discrepancy shows that improvement of our ansatz Eq.~(\ref{eq20}) for the contact mictrostructure should be a priority for future refinement of our constitutive model.

The oscillations of the coarse-grained microstructure tensor $\langle\boldsymbol{nn}\rangle$, 
on the other hand, are not enslaved to those of $\boldsymbol{E}$. 
As a result, the phase angle for the coarse-grained microstructure $\nn$ 
evolves in better qualitative agreement with 
the contact microstructure $\langle\boldsymbol{nn}\rangle_{c}$ found from the DEM simulations. 
Figure~\ref{figure6} shows that with increasing $\dot{\gamma}_{\perp}$,
the corresponding phase angle 
transitions smoothly
from $\theta\approx \pi/2$ (anti-phase with the $zy$-shear rate) 
to $\theta\approx 0$ (anti-phase with the $zy$-strain). 
The transition in the constitutive model 
occurs when the oscillation frequency $\omega$ 
exceeds the microstructure formation rate $\dot{\gamma}\beta/50$ (see Fig. \ref{figure2}b),
which corresponds to $\dot{\gamma}_{\perp}=\beta\gamma/50=10^{-2}$.
In the classification of~\cite{lin2016tunable}, the model seemingly captures both regime (i), instant adaptation, and regime (iii) chain breaking. However, the peak in the phase angle plot at $\dot{\gamma}_\perp\approx1$, which corresponds to the chain-tilting regime (ii),  is notably absent from the prediction of our constitutive model. This regime is characterized by a pseudo-elastic microstructural response while the stress response itself remains viscous. (See Section \ref{conclusions} for a further discussion.)
\subsection{Role of $\dot{\gamma}_\perp$: time-averaged response}

\begin{figure}
\includegraphics[trim = 0mm 0mm 0mm 0mm,clip,width=0.43\linewidth,page=1]{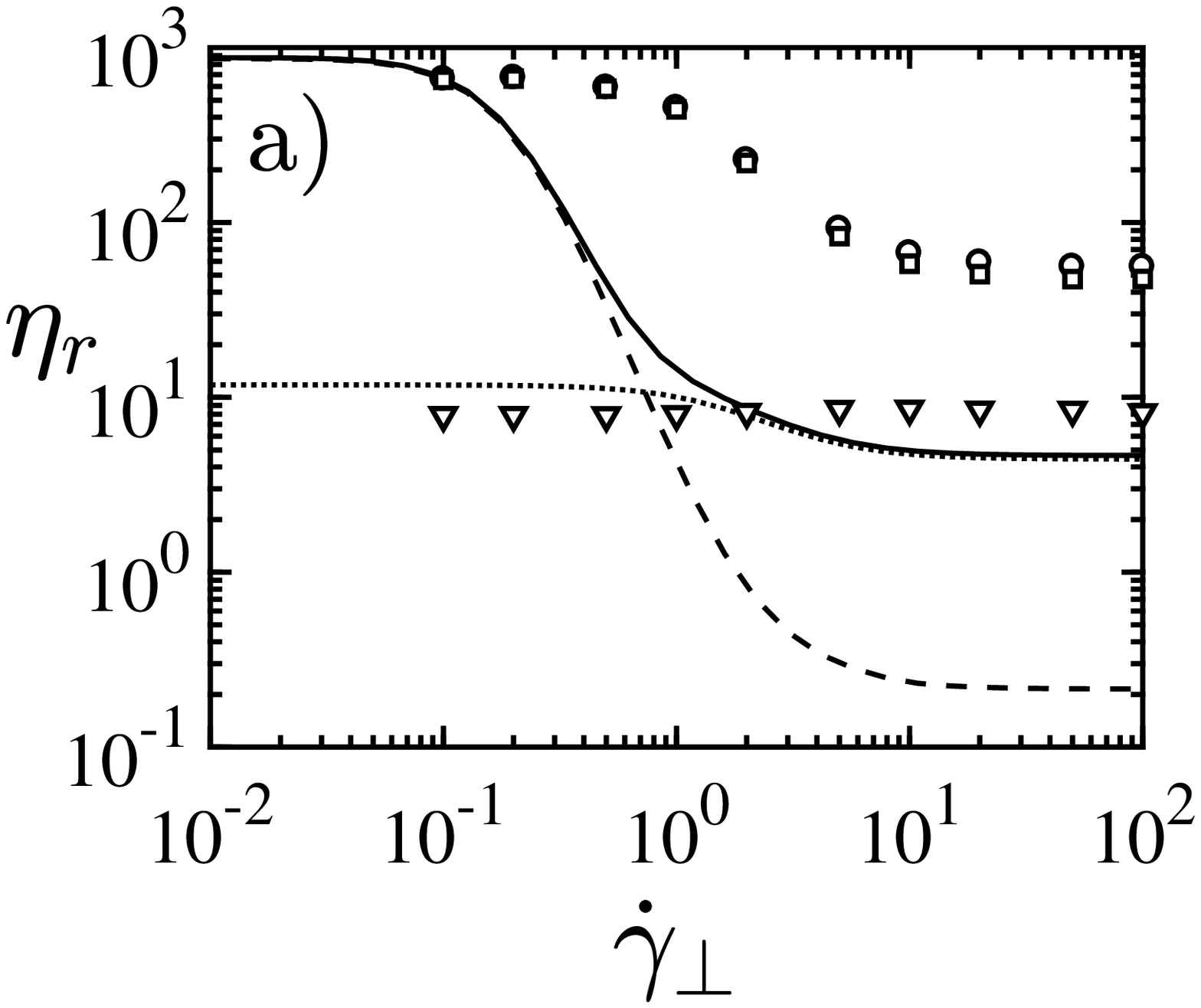}
\includegraphics[trim = 0mm 0mm 0mm 0mm,clip,width=0.47\linewidth,page=1]{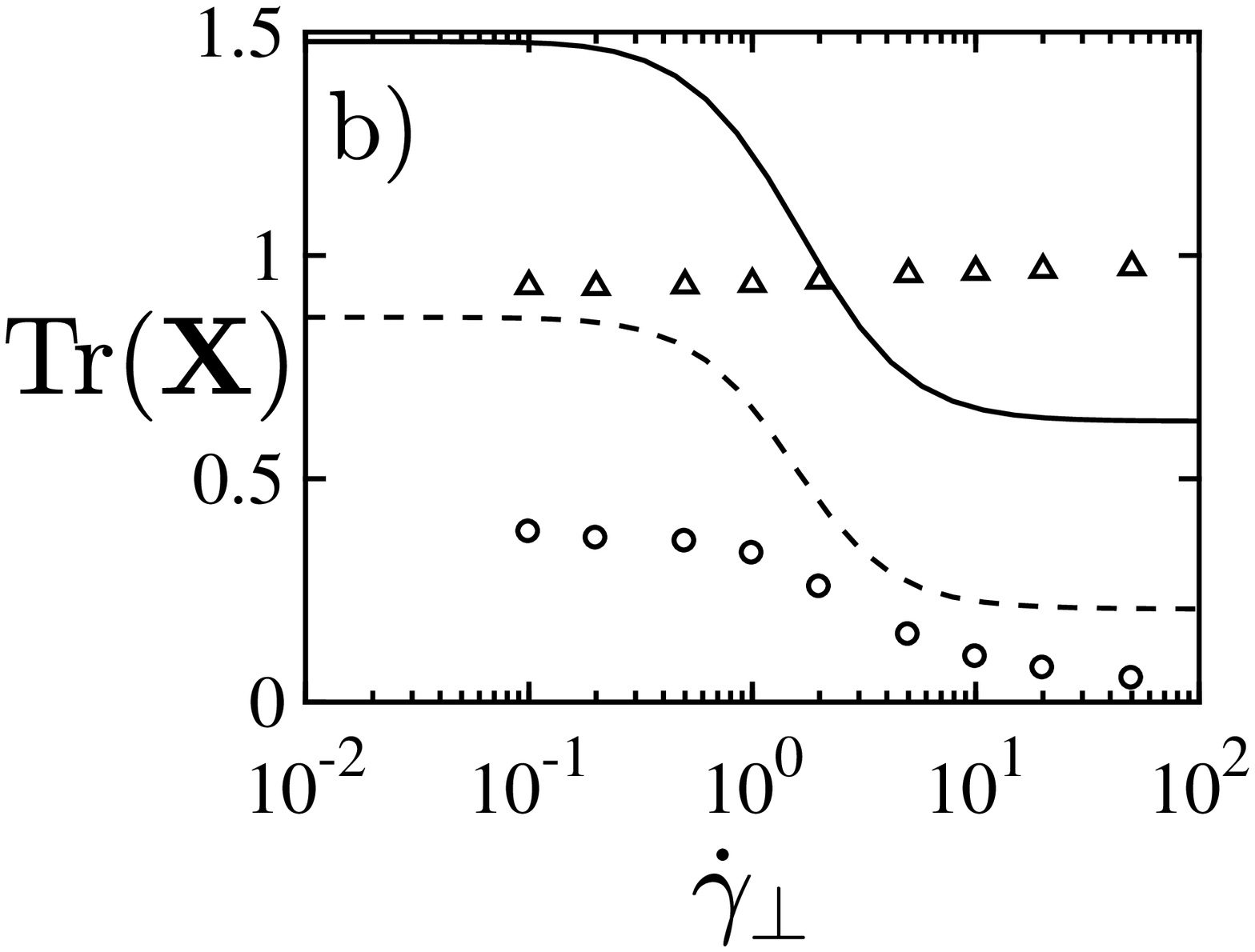}
\includegraphics[trim = 0mm 0mm 0mm 0mm,clip,width=0.45\linewidth,page=1]{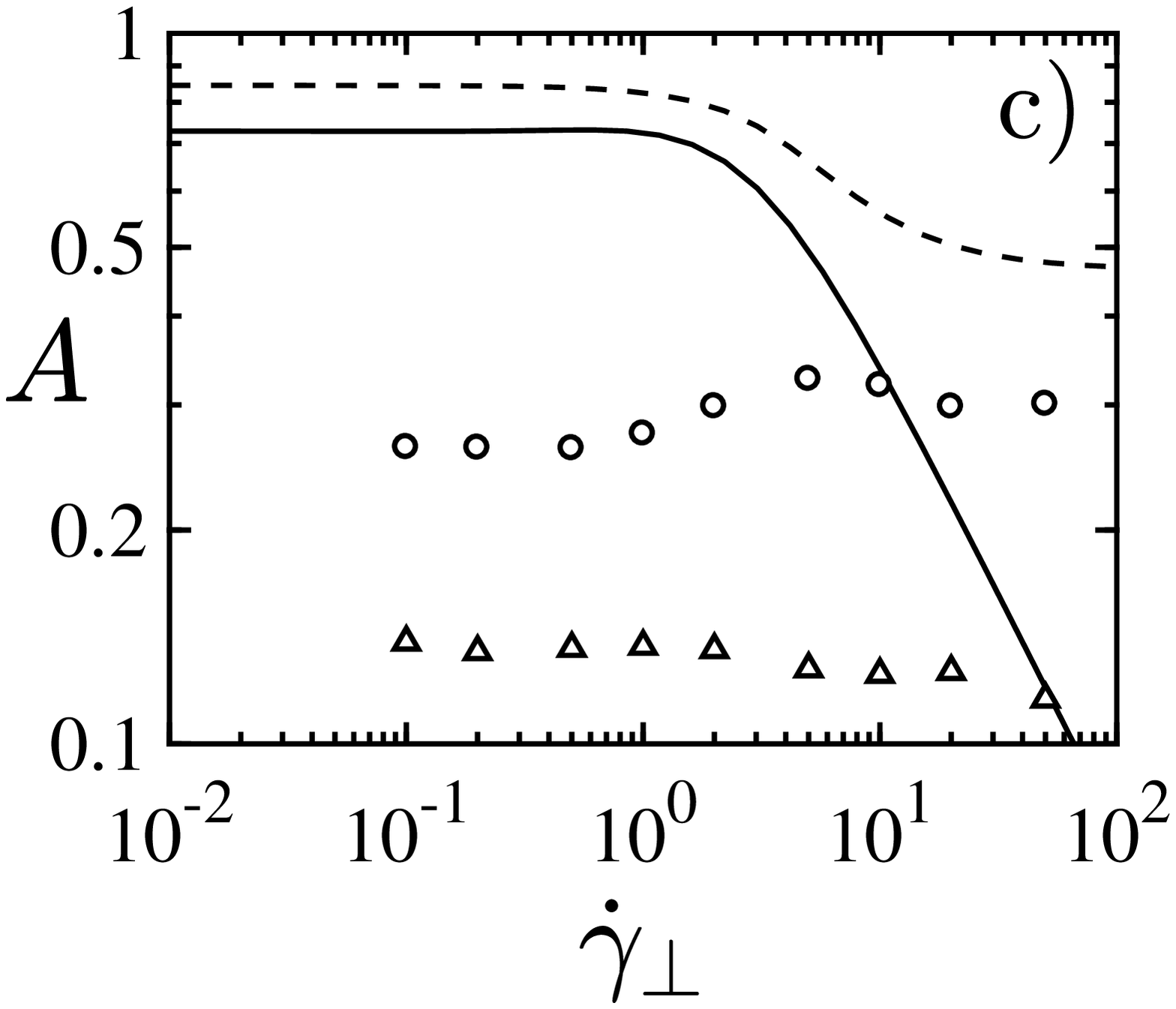}
\caption{
Time-averaged response of constitutive model and DEM simulation to
shear flow with superposed transverse oscillations.
(a) Total suspension viscosity $\Sigma_{xy}/\dot{\gamma}\eta_s$ (circles and solid line),
contact contribution to viscosity (squares and dashed line) and
hydrodynamic contribution to viscosity (triangles and dotted line)
in constitutive model (lines) and DEM (markers).
(b) Number of coarse-grained interactions 
$\mathrm{Tr}\langle\boldsymbol{nn}\rangle$ (triangles and solid line)
and contact interactions 
$\mathrm{Tr}\langle\boldsymbol{nn}\rangle_c$  (circles and dashed line)
normalised by $\xi^J_1$ [Eq. (\ref{eq34})] in constitutive model (lines),
and normalised by $Z^J_1=6$ in DEM (markers).
(c)  Anisotropy 
$A$ [Eq. (\ref{eq50})] of the 
coarse-grained microstructure $\boldsymbol{X} = \nn$ (triangles and solid line) and 
contact microstructure $\boldsymbol{X} = \nn_c$ (circles and dashed line),
in constitutive model (lines) and DEM (markers).
}
\label{figure4}
\end{figure}

We next present the viscosity and the microstructure, as
predicted by the DEM simulation and the constitutive model,
averaged over the oscillation cycle,
again focussing on the fully shear-thickened case with $\dot{\gamma}_r=\infty$.

The constitutive model qualitatively predicts the $\dot{\gamma}_\perp$-mediated 
decrease in suspension viscosity $\eta_r=\Sigma_{xy}/(\dot{\gamma}\eta_s)$, Fig.~\ref{figure4}a,
consistent with our DEM simulation data and with experimental data measured under equivalent shearing conditions~\cite{lin2016tunable} and indeed
under acoustic perturbations~\cite{sehgal2019using}.  
Quantitatively, however, the DEM simulations show a decrease in the viscosity by a factor of around 20 at this volume fraction (in an earlier article we showed within DEM the dependence of this decrease on $\phi$~\cite{ness2018shaken}), 
whereas the constitutive model predicts a drop by a factor of around 200.

This difference reflects that
the contact stress in the constitutive model depends too strongly on the number of contacts [Eq.~(\ref{eq1c})].
The main discrepancy is that while the constitutive model predicts for large $\dot\gamma_\perp$ a complete collapse of the contact contribution leaving only the lubrication part, the DEM data shows that the stress remains contact-dominated even at large $\dot\gamma_\perp$. Although in this regime (the chain-breaking regime of~\cite{lin2016tunable}) the microstructure is severely disrupted, in the DEM simulations direct contacts are not so diminished as to contribute negligibly to stress, as the constitutive model predicts. This is due in part to the chosen operating condition of $\phi=0.56$ and $\mu=1$. Close to $\phi_2^J$, even small numbers of frictional contacts are sufficient to give a dominant contact stress. At lower $\phi$ and $\mu$ the DEM simulation does indeed predict hydrodynamic stress dominance at large $\dot{\gamma}_\perp$~\cite{ness2018shaken}.

Shown in Fig.~\ref{figure4}b 
are the number of contact interactions $\mathrm{Tr}\langle\boldsymbol{nn}\rangle_c$ 
and the number of coarse-grained interactions $\mathrm{Tr}\langle\boldsymbol{nn}\rangle$,
as functions of $\dot{\gamma}_\perp$ predicted by the DEM simulation and the constitutive model.
For the DEM simulations, $\langle\boldsymbol{nn}\rangle_c$ includes 
direct contacts only [those for which we compute Eq.~(\ref{cundall})], 
while 
$\langle\boldsymbol{nn}\rangle$ corresponds to 
all (direct and lubricated) interactions 
within the lubrication cutoff length $h=0.05 a$.
$\mathrm{Tr}\langle\boldsymbol{nn}\rangle_c$ decreases steadily with increasing $\dot{\gamma}_{\perp}$
as the 
oscillations increasingly break up force chains created by the steady shearing flow.
$\mathrm{Tr}\langle\boldsymbol{nn}\rangle$, 
on the other hand, is only weakly affected by the transverse oscillations.
This is due to the low strain amplitude, $\gamma=0.01$, which is sufficient to move particles out of direct contact,
but not to move interacting particles out of each other's lubrication films 
(as cut off at $h=0.05 a$).
This result is independent of $\gamma$, provided $\gamma$ remains within the range mentioned above (approximately $10^{-4}<\gamma<0.05$).
Overall there is qualitative agreement in $\mathrm{Tr}\langle\boldsymbol{nn}\rangle_c$ 
between the constitutive model and discrete-element simulation,
with both predicting a similar 
$\dot{\gamma}_{\perp}$-dependence.
For $\mathrm{Tr}\langle\boldsymbol{nn}\rangle$ on the other hand, there is disagreement,
where the constitutive model predicts a decrease, and 
the DEM predicts a constant as a function of $\dot\gamma_\perp$.

Fig.~\ref{figure4}c shows the time-averaged microstructural anisotropy, defined as
\begin{equation} 
A=\left\{1-27\mathrm{Det}\left[\boldsymbol{X}/\mathrm{Tr}(\boldsymbol{X})\right]\right\}^{\frac{1}{3}}, 
\label{eq50}
\end{equation}
for the contact microstructure 
$\boldsymbol{X}=\langle\boldsymbol{nn}\rangle_c$ and for the coarse-grained microstructure 
$\boldsymbol{X}=\langle\boldsymbol{nn}\rangle$, both in the constitutive model and in DEM.
In DEM $\langle\boldsymbol{nn}\rangle_c$ has 
$A\approx 0.3$
and $\langle\boldsymbol{nn}\rangle$ has 
${A}\approx 0.1$. 
In the constitutive model the microstructure is more anisotropic with 
${A}\approx 0.8$, 
for $\langle\boldsymbol{nn}\rangle_c$ and 
${A}\approx 0.7$, 
for $\langle\boldsymbol{nn}\rangle$, 
until $\dot\gamma_\perp$ becomes large. 
This overprediction of microstructural anisotropy within our constitutive model has previously been reported for reversal flows~\cite{gillissen2019constitutive}. 

When $\dot\gamma_\perp$ does become large, the constitutive model predicts 
near isotropization of the coarse-grained microstructure, 
$A= \mathcal{O}(\beta\gamma/50)$,
and saturating anisotropy of the contact microstructure, 
$A\approx 0.5\phi$,
which is found by inserting 
$\tfrac{1}{2}\left(\boldsymbol{\delta}_y\boldsymbol{\delta}_y
+\boldsymbol{\delta}_z\boldsymbol{\delta}_z\right)$ 
for the time averaged value for
$-\boldsymbol{E}_c/\vert\boldsymbol{E}_c\vert$, 
and $\langle\boldsymbol{nn}\rangle\approx\phi\boldsymbol{\delta}/3$ into 
 Eqs.~(\ref{eq13}, \ref{eq20}, \ref{eq50}).
 This contrasts with the DEM behaviour which shows a mild maximum in anisotropy in $\langle\boldsymbol{nn}\rangle_c$ in the chain-tilting regime (regime (ii) as defined above). 
This discrepancy is presumably related to the failure to capture the phase angle between the $yz$-component of $\langle\boldsymbol{nn}\rangle_c$ and the transverse strain in this regime (see Fig.~\ref{figure6}).
Failure of the DEM to reach complete isotropy in $\langle\boldsymbol{nn}\rangle$ and $\langle\boldsymbol{nn}\rangle_c$ at large $\dot{\gamma}_\perp$
is consistent with there being continuing dominance of the contact contribution to the shear stress, discussed above.

\begin{figure}
\centerline{
\psfig{file=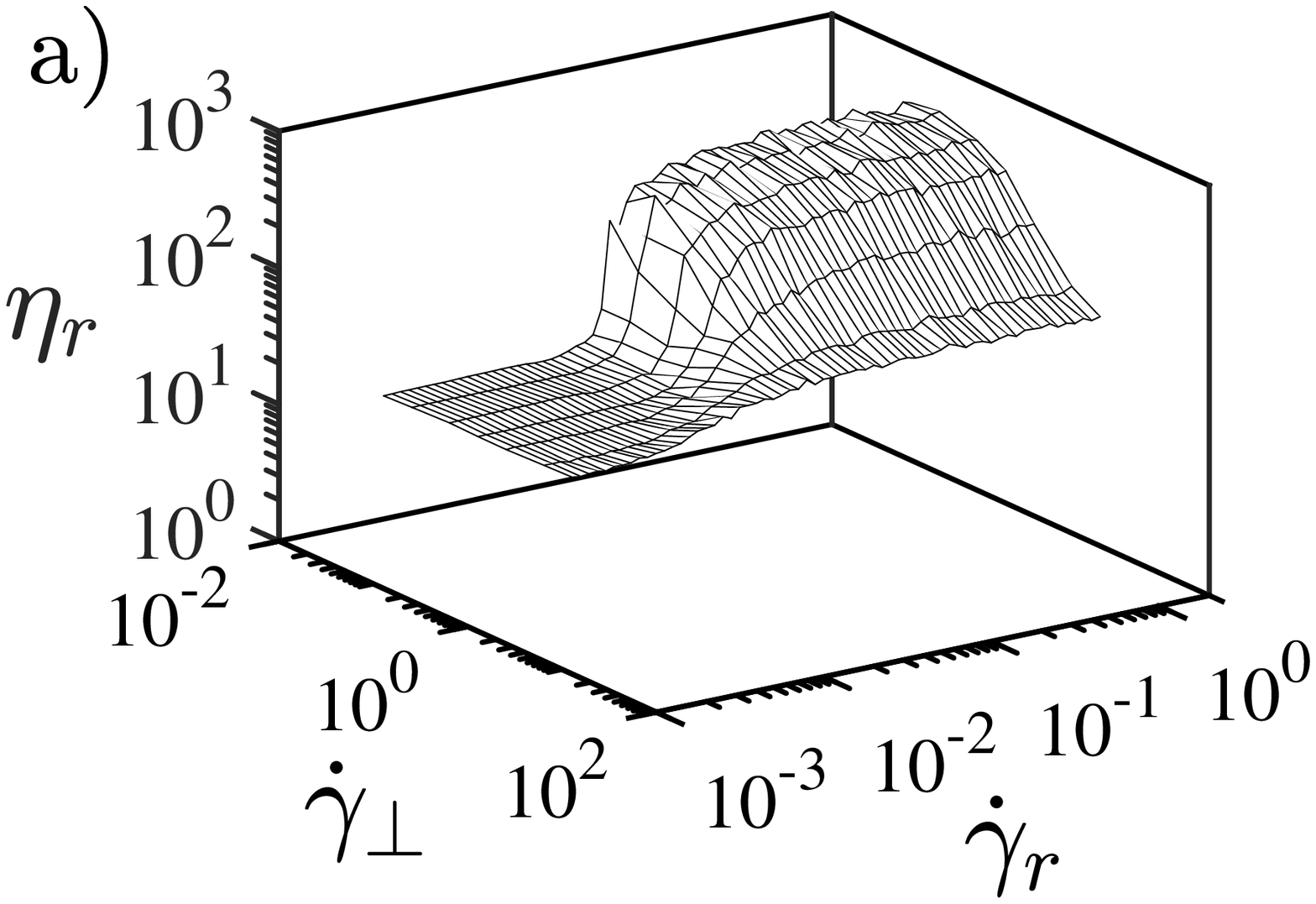,width=0.5\linewidth}
\psfig{file=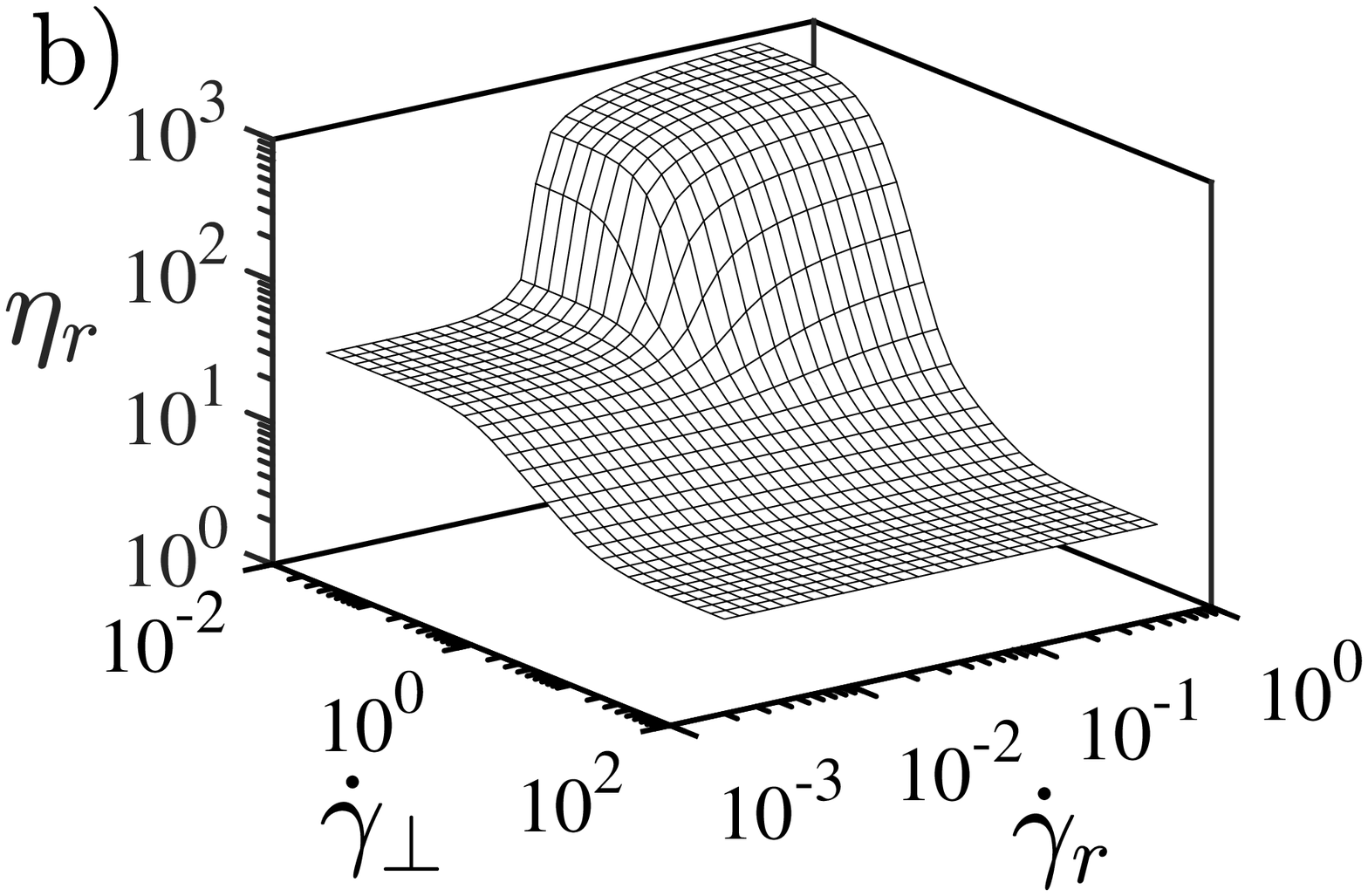,width=0.5\linewidth}
}
\caption{
Suspension viscosity as a function of 
dimensionless shear rate $\dot{\gamma}_r$ and dimensionless frequency $\dot{\gamma}_{\perp}$
for DEM (a) and constitutive model (b).
}\label{figure5}
\end{figure}

We finally present in Fig.~\ref{figure5} 
plots of the suspension viscosity
as functions of $\dot{\gamma}_r$ and $\dot{\gamma}_\perp$, found by DEM simulation and predicted by the constitutive model (the simulated large $\dot{\gamma}_r$ data were previously reported in Ref.~\cite{ness2018shaken}).
The model and DEM simulation both predict that the viscosity reduction obtained under transverse oscillatory shearing is largest
for shear-thickened suspensions.
This follows naturally from the fact that the oscillations act by breaking up particle-particle contacts: frictional flowing states of $\Pi \gg \Pi^*$ are dominated by particle-particle contact stresses and stand to lose a substantially larger proportion of their viscosity by having such contacts removed, compared to lubrication-dominated suspensions ($\Pi\ll\Pi^*$). For these purposes the term `shear-thickened' suspensions of course include rate-independent materials of high friction for which $\Pi^*$ is effectively zero~\cite{guy2015towards}.

Overall the qualitative agreement between Figs.~\ref{figure5}a, b represents encouraging success of our constitutive model under conditions of both rate- and time- dependent flow. Nonetheless, some discrepancies are apparent
within the ($\dot\gamma_r,\dot\gamma_\perp$) range shown here. 
At small $\dot\gamma_r$,
for which contacts are frictionless and the resulting DEM contact stress is subdominant,
the viscosity in DEM simulation is roughly independent of $\dot\gamma_\perp$.
Under these conditions, where the hydrodynamic stress is dominant, it is to be expected that the oscillation-mediated loss of contacts does not lead to a significant change in the viscosity.
The constitutive model, meanwhile, predicts a decrease in viscosity with $\dot{\gamma}_\perp$ at small $\dot{\gamma}_r$.
This reflects that 
changes in the coarse-grained microstructure and the lubrication stress are more pronounced in the constitutive model than in the DEM.
At large $\dot\gamma_\perp$, the viscosity in DEM 
increases with $\dot\gamma_r$
since (at this proximity to $\phi_2^J$)
the onset of friction leads to a substantial contact stress
(albeit lower than when $\dot{\gamma}_\perp$ is small).
In the constitutive model, however, the viscosity at large $\dot{\gamma}_\perp$ is independent of $\dot{\gamma}_r$, reflecting that 
the modelled oscillations 
over-predict the break-up of the microstructure (Fig. \ref{figure4}),
and providing further indication that the contact stress is too sensitive to the number of contacts [Eq.~(\ref{eq19})].

\section{Discussion and Conclusions}
\label{conclusions}
We have presented a self-contained derivation for a recently proposed constitutive model for the microstructure and stress of shear-thickening particle suspensions, discussing {\em en route} the roles played by the `jamming coordinate'  $\xi=\mathrm{Tr}\nn_c$ and the contact microstructure $\nn_c$, which is related, within the model, to  
the coarse-grained microstructure tensor $\nn$. 
This relation allows a closed constitutive model at the coarse-grained level,
whilst making testable predictions for the contact statistics. 
Along with the stress and other observable quantities, these can be compared with experiment or, as done in this paper, with simulations of particle-based models based on the DEM (discrete element method) formalism. The use of DEM simulation data, for which microstructural data can be interrogated almost {\em ad infinitum} (in contrast to experiments which generally cannot resolve individual contact forces), offers a set of stringent tests for rheological constitutive models, as emphasized recently by Chacko et al.~\cite{chacko2018shear}.

In Ref.~\cite{gillissen2019constitutive} we confronted the new constitutive model with such data for the case of reversal of steady shear flow. The model was found qualitatively correct in most aspects, but with a systematic over-prediction of microstructural anisotropy which was reflected in relatively poor prediction of normal stress differences. 

In the present paper we have taken the simulation-based testing of the model considerably further, by addressing steady shear flows with superposed transverse oscillations. For friction-dominated systems ($\Pi\gg\Pi^*$) this protocol has been shown capable of drastically reducing the mean viscosity, in some cases unjamming systems whose viscosity would otherwise be infinite~\cite{lin2016tunable,ness2018shaken}. 
This protocol may find utility in active rheology control~\cite{sehgal2019using} for various industrial applications. Because of its strong influence on time-dependent suspension microstructure, it provides a range of stringent tests for any constitutive model. Since flow conditions evolve continuously, these tests complement those offered by sudden flow reversal.

Overall we again found qualitative agreement between the constitutive model predictions and data generated by discrete-element simulation. This applies in particular to the 
decrease in the contacts with increasing oscillation frequency (Fig. \ref{figure4}b),
and to the transient build-up of the amplitude of the transverse
shear stress and its phase relative to the transverse shear rate (Figs. \ref{figure3}a, b).

However, the model falls short in other respects, such as the corresponding phase relation for the 
$yz$-component of the contact microstructure $\nn_c$.  This shows a discrepancy that is at least partly the fault of our ansatz for 
$\nn_c$ in terms of $\nn$, given in Eq.~(\ref{eq20}). Improvement to this ansatz is therefore a target for 
future refinement of our constitutive model.
A second discrepancy is that the constitutive model predicts the contact contribution to viscosity to collapse to extremely low levels at high transverse oscillation frequencies, so that lubrication terms dominate, whereas the DEM simulations show the collapse to be much more moderate, with direct contact terms still dominating the stress, at least when $\phi$ is close to the frictional jamming point $\phi_2^J$. 
Thirdly, although Fig.~\ref{figure5} shows broad qualitative agreement for the viscosity as a function of the reduced shear rate $\dot\gamma_r$ and oscillation frequency $\dot\gamma_\perp$, the behaviours seen in DEM simulations along both the small $\dot\gamma_r$ and the large $\dot\gamma_\perp$ edges of the diagram are not properly captured by the constitutive model. 

The explanations of these shortcomings remain a topic of ongoing research, to which we hope to return in future publications. Candidates for improvement include not only the specific approximation for $\nn_c$ mentioned above [Eq. (\ref{eq20})], but also 
the relation between the contact force and the number of contacts [$C_3$-term in Eq.~(\ref{eq1})];
a microstructure-based interaction force with the background [$C_1$-term in Eq.~(\ref{eq1})], 
which should limit anisotropy in dense systems; 
the assumption of a friction-independent microstructural evolution [absence of a friction term in Eq.~(\ref{eq1})]; 
the Hinch-Leal type closure relation [Eq. (\ref{eq13})]; 
and our simplified approach to the angular distribution of birth and death processes among contacts [Eq. (\ref{eq9})].

Pending further exploration of all these aspects, the work reported above already confirms the value of comparing constitutive models for dense suspensions not only with macroscopic experimental observations (which are generally limited to measurements of stress), but also with particle-based simulations that can give detailed microstructural statistics. Such comparisons increasingly allow the assumptions of the model to be tested individually rather than collectively, an approach that we hope should speed future progress towards a fully predictive constitutive rheology for dense suspensions, both shear-thickening and otherwise.

{\em Acknowledgements}: We acknowledge financial support from the Engineering and Physical Sciences Research Council of the United Kingdom Grant No. EP/N024915/1, and from the European Research Council under the Horizon 2020 Programme, ERC grant agreement number 740269. MEC is funded by the Royal Society. CN is funded by the Maudslay-Butler Research Fellowship at Pembroke College, Cambridge.

\bibliography{article3}

\begin{thebibliography}{34}%
\makeatletter
\providecommand \@ifxundefined [1]{%
 \@ifx{#1\undefined}
}%
\providecommand \@ifnum [1]{%
 \ifnum #1\expandafter \@firstoftwo
 \else \expandafter \@secondoftwo
 \fi
}%
\providecommand \@ifx [1]{%
 \ifx #1\expandafter \@firstoftwo
 \else \expandafter \@secondoftwo
 \fi
}%
\providecommand \natexlab [1]{#1}%
\providecommand \enquote  [1]{``#1''}%
\providecommand \bibnamefont  [1]{#1}%
\providecommand \bibfnamefont [1]{#1}%
\providecommand \citenamefont [1]{#1}%
\providecommand \href@noop [0]{\@secondoftwo}%
\providecommand \href [0]{\begingroup \@sanitize@url \@href}%
\providecommand \@href[1]{\@@startlink{#1}\@@href}%
\providecommand \@@href[1]{\endgroup#1\@@endlink}%
\providecommand \@sanitize@url [0]{\catcode `\\12\catcode `\$12\catcode
  `\&12\catcode `\#12\catcode `\^12\catcode `\_12\catcode `\%12\relax}%
\providecommand \@@startlink[1]{}%
\providecommand \@@endlink[0]{}%
\providecommand \url  [0]{\begingroup\@sanitize@url \@url }%
\providecommand \@url [1]{\endgroup\@href {#1}{\urlprefix }}%
\providecommand \urlprefix  [0]{URL }%
\providecommand \Eprint [0]{\href }%
\providecommand \doibase [0]{http://dx.doi.org/}%
\providecommand \selectlanguage [0]{\@gobble}%
\providecommand \bibinfo  [0]{\@secondoftwo}%
\providecommand \bibfield  [0]{\@secondoftwo}%
\providecommand \translation [1]{[#1]}%
\providecommand \BibitemOpen [0]{}%
\providecommand \bibitemStop [0]{}%
\providecommand \bibitemNoStop [0]{.\EOS\space}%
\providecommand \EOS [0]{\spacefactor3000\relax}%
\providecommand \BibitemShut  [1]{\csname bibitem#1\endcsname}%
\let\auto@bib@innerbib\@empty
\bibitem [{\citenamefont {Mari}\ \emph {et~al.}(2014)\citenamefont {Mari},
  \citenamefont {Seto}, \citenamefont {Morris},\ and\ \citenamefont
  {Denn}}]{mari2014shear}%
  \BibitemOpen
  \bibfield  {author} {\bibinfo {author} {\bibfnamefont {Romain}\ \bibnamefont
  {Mari}}, \bibinfo {author} {\bibfnamefont {Ryohei}\ \bibnamefont {Seto}},
  \bibinfo {author} {\bibfnamefont {Jeffrey~F}\ \bibnamefont {Morris}}, \ and\
  \bibinfo {author} {\bibfnamefont {Morton~M}\ \bibnamefont {Denn}},\
  }\bibfield  {title} {\enquote {\bibinfo {title} {Shear thickening,
  frictionless and frictional rheologies in non-brownian suspensions},}\
  }\href@noop {} {\bibfield  {journal} {\bibinfo  {journal} {J. Rheol.}\
  }\textbf {\bibinfo {volume} {58}},\ \bibinfo {pages} {1693--1724} (\bibinfo
  {year} {2014})}\BibitemShut {NoStop}%
\bibitem [{\citenamefont {Seto}\ \emph {et~al.}(2013)\citenamefont {Seto},
  \citenamefont {Mari}, \citenamefont {Morris},\ and\ \citenamefont
  {Denn}}]{seto2013discontinuous}%
  \BibitemOpen
  \bibfield  {author} {\bibinfo {author} {\bibfnamefont {Ryohei}\ \bibnamefont
  {Seto}}, \bibinfo {author} {\bibfnamefont {Romain}\ \bibnamefont {Mari}},
  \bibinfo {author} {\bibfnamefont {Jeffrey~F}\ \bibnamefont {Morris}}, \ and\
  \bibinfo {author} {\bibfnamefont {Morton~M}\ \bibnamefont {Denn}},\
  }\bibfield  {title} {\enquote {\bibinfo {title} {Discontinuous shear
  thickening of frictional hard-sphere suspensions},}\ }\href@noop {}
  {\bibfield  {journal} {\bibinfo  {journal} {Phys. Rev. Lett.}\ }\textbf
  {\bibinfo {volume} {111}},\ \bibinfo {pages} {218301} (\bibinfo {year}
  {2013})}\BibitemShut {NoStop}%
\bibitem [{\citenamefont {Guy}\ \emph {et~al.}(2015)\citenamefont {Guy},
  \citenamefont {Hermes},\ and\ \citenamefont {Poon}}]{guy2015towards}%
  \BibitemOpen
  \bibfield  {author} {\bibinfo {author} {\bibfnamefont {B.~M.}\ \bibnamefont
  {Guy}}, \bibinfo {author} {\bibfnamefont {Michiel}\ \bibnamefont {Hermes}}, \
  and\ \bibinfo {author} {\bibfnamefont {Wilson C.~K.}\ \bibnamefont {Poon}},\
  }\bibfield  {title} {\enquote {\bibinfo {title} {Towards a unified
  description of the rheology of hard-particle suspensions},}\ }\href@noop {}
  {\bibfield  {journal} {\bibinfo  {journal} {Phys. Rev. Lett.}\ }\textbf
  {\bibinfo {volume} {115}},\ \bibinfo {pages} {088304} (\bibinfo {year}
  {2015})}\BibitemShut {NoStop}%
\bibitem [{\citenamefont {Lin}\ \emph {et~al.}(2015)\citenamefont {Lin},
  \citenamefont {Guy}, \citenamefont {Hermes}, \citenamefont {Ness},
  \citenamefont {Sun}, \citenamefont {Poon},\ and\ \citenamefont
  {Cohen}}]{lin2015hydrodynamic}%
  \BibitemOpen
  \bibfield  {author} {\bibinfo {author} {\bibfnamefont {Neil Y.~C.}\
  \bibnamefont {Lin}}, \bibinfo {author} {\bibfnamefont {Ben~M}\ \bibnamefont
  {Guy}}, \bibinfo {author} {\bibfnamefont {Michiel}\ \bibnamefont {Hermes}},
  \bibinfo {author} {\bibfnamefont {Chris}\ \bibnamefont {Ness}}, \bibinfo
  {author} {\bibfnamefont {Jin}\ \bibnamefont {Sun}}, \bibinfo {author}
  {\bibfnamefont {Wilson C.~K.}\ \bibnamefont {Poon}}, \ and\ \bibinfo {author}
  {\bibfnamefont {Itai}\ \bibnamefont {Cohen}},\ }\bibfield  {title} {\enquote
  {\bibinfo {title} {Hydrodynamic and contact contributions to continuous shear
  thickening in colloidal suspensions},}\ }\href@noop {} {\bibfield  {journal}
  {\bibinfo  {journal} {Phys. Rev. Lett.}\ }\textbf {\bibinfo {volume} {115}},\
  \bibinfo {pages} {228304} (\bibinfo {year} {2015})}\BibitemShut {NoStop}%
\bibitem [{\citenamefont {Royer}\ \emph {et~al.}(2016)\citenamefont {Royer},
  \citenamefont {Blair},\ and\ \citenamefont {Hudson}}]{royer2016rheological}%
  \BibitemOpen
  \bibfield  {author} {\bibinfo {author} {\bibfnamefont {J~R}\ \bibnamefont
  {Royer}}, \bibinfo {author} {\bibfnamefont {Dl~L}\ \bibnamefont {Blair}}, \
  and\ \bibinfo {author} {\bibfnamefont {S~D}\ \bibnamefont {Hudson}},\
  }\bibfield  {title} {\enquote {\bibinfo {title} {Rheological signature of
  frictional interactions in shear thickening suspensions},}\ }\href@noop {}
  {\bibfield  {journal} {\bibinfo  {journal} {Phys. Rev. Lett.}\ }\textbf
  {\bibinfo {volume} {116}},\ \bibinfo {pages} {188301} (\bibinfo {year}
  {2016})}\BibitemShut {NoStop}%
\bibitem [{\citenamefont {Wagner}\ and\ \citenamefont
  {Brady}(2009)}]{wagner2009shear}%
  \BibitemOpen
  \bibfield  {author} {\bibinfo {author} {\bibfnamefont {Norman~J}\
  \bibnamefont {Wagner}}\ and\ \bibinfo {author} {\bibfnamefont {John~F}\
  \bibnamefont {Brady}},\ }\bibfield  {title} {\enquote {\bibinfo {title}
  {Shear thickening in colloidal dispersions},}\ }\href@noop {} {\bibfield
  {journal} {\bibinfo  {journal} {Phys. Today}\ }\textbf {\bibinfo {volume}
  {62}},\ \bibinfo {pages} {27--32} (\bibinfo {year} {2009})}\BibitemShut
  {NoStop}%
\bibitem [{\citenamefont {Jamali}\ and\ \citenamefont
  {Brady}(2019)}]{jamali2019alternative}%
  \BibitemOpen
  \bibfield  {author} {\bibinfo {author} {\bibfnamefont {S}~\bibnamefont
  {Jamali}}\ and\ \bibinfo {author} {\bibfnamefont {J~F}\ \bibnamefont
  {Brady}},\ }\bibfield  {title} {\enquote {\bibinfo {title} {Alternative
  frictional model for discontinuous shear thickening of dense suspensions:
  Hydrodynamics},}\ }\href@noop {} {\bibfield  {journal} {\bibinfo  {journal}
  {Phys. Rev. Lett.}\ }\textbf {\bibinfo {volume} {123}},\ \bibinfo {pages}
  {138002} (\bibinfo {year} {2019})}\BibitemShut {NoStop}%
\bibitem [{\citenamefont {Wyart}\ and\ \citenamefont
  {Cates}(2014)}]{wyart2014discontinuous}%
  \BibitemOpen
  \bibfield  {author} {\bibinfo {author} {\bibfnamefont {Matthieu}\
  \bibnamefont {Wyart}}\ and\ \bibinfo {author} {\bibfnamefont {M.~E.}\
  \bibnamefont {Cates}},\ }\bibfield  {title} {\enquote {\bibinfo {title}
  {Discontinuous shear thickening without inertia in dense non-brownian
  suspensions},}\ }\href@noop {} {\bibfield  {journal} {\bibinfo  {journal}
  {Phys. Rev. Lett.}\ }\textbf {\bibinfo {volume} {112}},\ \bibinfo {pages}
  {098302} (\bibinfo {year} {2014})}\BibitemShut {NoStop}%
\bibitem [{\citenamefont {Boyer}\ \emph {et~al.}(2011)\citenamefont {Boyer},
  \citenamefont {Guazzelli},\ and\ \citenamefont
  {Pouliquen}}]{boyer2011unifying}%
  \BibitemOpen
  \bibfield  {author} {\bibinfo {author} {\bibfnamefont {Fran{\c{c}}ois}\
  \bibnamefont {Boyer}}, \bibinfo {author} {\bibfnamefont {{\'E}lisabeth}\
  \bibnamefont {Guazzelli}}, \ and\ \bibinfo {author} {\bibfnamefont {Olivier}\
  \bibnamefont {Pouliquen}},\ }\bibfield  {title} {\enquote {\bibinfo {title}
  {Unifying suspension and granular rheology},}\ }\href@noop {} {\bibfield
  {journal} {\bibinfo  {journal} {Phys. Rev. Lett.}\ }\textbf {\bibinfo
  {volume} {107}},\ \bibinfo {pages} {188301} (\bibinfo {year}
  {2011})}\BibitemShut {NoStop}%
\bibitem [{\citenamefont {Comtet}\ \emph {et~al.}(2017)\citenamefont {Comtet},
  \citenamefont {Chatt{\'e}}, \citenamefont {Nigu{\`e}s}, \citenamefont
  {Bocquet}, \citenamefont {Siria},\ and\ \citenamefont
  {Colin}}]{comtet2017pairwise}%
  \BibitemOpen
  \bibfield  {author} {\bibinfo {author} {\bibfnamefont {Jean}\ \bibnamefont
  {Comtet}}, \bibinfo {author} {\bibfnamefont {Guillaume}\ \bibnamefont
  {Chatt{\'e}}}, \bibinfo {author} {\bibfnamefont {Antoine}\ \bibnamefont
  {Nigu{\`e}s}}, \bibinfo {author} {\bibfnamefont {Lyd{\'e}ric}\ \bibnamefont
  {Bocquet}}, \bibinfo {author} {\bibfnamefont {Alessandro}\ \bibnamefont
  {Siria}}, \ and\ \bibinfo {author} {\bibfnamefont {Annie}\ \bibnamefont
  {Colin}},\ }\bibfield  {title} {\enquote {\bibinfo {title} {Pairwise
  frictional profile between particles determines discontinuous shear
  thickening transition in non-colloidal suspensions},}\ }\href@noop {}
  {\bibfield  {journal} {\bibinfo  {journal} {Nat. Commun.}\ }\textbf {\bibinfo
  {volume} {8}},\ \bibinfo {pages} {15633} (\bibinfo {year}
  {2017})}\BibitemShut {NoStop}%
\bibitem [{\citenamefont {Krieger}\ and\ \citenamefont
  {Dougherty}(1959)}]{krieger1959mechanism}%
  \BibitemOpen
  \bibfield  {author} {\bibinfo {author} {\bibfnamefont {Irvin~M}\ \bibnamefont
  {Krieger}}\ and\ \bibinfo {author} {\bibfnamefont {Thomas~J}\ \bibnamefont
  {Dougherty}},\ }\bibfield  {title} {\enquote {\bibinfo {title} {A mechanism
  for non-newtonian flow in suspensions of rigid spheres},}\ }\href@noop {}
  {\bibfield  {journal} {\bibinfo  {journal} {Trans. Soc. Rheol.}\ }\textbf
  {\bibinfo {volume} {3}},\ \bibinfo {pages} {137--152} (\bibinfo {year}
  {1959})}\BibitemShut {NoStop}%
\bibitem [{\citenamefont {Hermes}\ \emph {et~al.}(2016)\citenamefont {Hermes},
  \citenamefont {Guy}, \citenamefont {Poon}, \citenamefont {Poy}, \citenamefont
  {Cates},\ and\ \citenamefont {Wyart}}]{hermes2016unsteady}%
  \BibitemOpen
  \bibfield  {author} {\bibinfo {author} {\bibfnamefont {Michiel}\ \bibnamefont
  {Hermes}}, \bibinfo {author} {\bibfnamefont {Ben~M}\ \bibnamefont {Guy}},
  \bibinfo {author} {\bibfnamefont {Wilson C.~K.}\ \bibnamefont {Poon}},
  \bibinfo {author} {\bibfnamefont {Guilhem}\ \bibnamefont {Poy}}, \bibinfo
  {author} {\bibfnamefont {Michael~E.}\ \bibnamefont {Cates}}, \ and\ \bibinfo
  {author} {\bibfnamefont {Matthieu}\ \bibnamefont {Wyart}},\ }\bibfield
  {title} {\enquote {\bibinfo {title} {Unsteady flow and particle migration in
  dense, non-brownian suspensions},}\ }\href@noop {} {\bibfield  {journal}
  {\bibinfo  {journal} {J. Rheol.}\ }\textbf {\bibinfo {volume} {60}},\
  \bibinfo {pages} {905--916} (\bibinfo {year} {2016})}\BibitemShut {NoStop}%
\bibitem [{\citenamefont {Guy}\ \emph {et~al.}(2020)\citenamefont {Guy},
  \citenamefont {Ness}, \citenamefont {Hermes}, \citenamefont {Sawiak},
  \citenamefont {Sun},\ and\ \citenamefont {Poon}}]{guy2020testing}%
  \BibitemOpen
  \bibfield  {author} {\bibinfo {author} {\bibfnamefont {Ben~M}\ \bibnamefont
  {Guy}}, \bibinfo {author} {\bibfnamefont {Christopher}\ \bibnamefont {Ness}},
  \bibinfo {author} {\bibfnamefont {Michiel}\ \bibnamefont {Hermes}}, \bibinfo
  {author} {\bibfnamefont {Laura~J}\ \bibnamefont {Sawiak}}, \bibinfo {author}
  {\bibfnamefont {Jin}\ \bibnamefont {Sun}}, \ and\ \bibinfo {author}
  {\bibfnamefont {Wilson~CK}\ \bibnamefont {Poon}},\ }\bibfield  {title}
  {\enquote {\bibinfo {title} {Testing the {W}yart--{C}ates model for
  non-{B}rownian shear thickening using bidisperse suspensions},}\ }\href@noop
  {} {\bibfield  {journal} {\bibinfo  {journal} {Soft Matter}\ }\textbf
  {\bibinfo {volume} {16}},\ \bibinfo {pages} {229--237} (\bibinfo {year}
  {2020})}\BibitemShut {NoStop}%
\bibitem [{\citenamefont {Gillissen}\ \emph {et~al.}(2019)\citenamefont
  {Gillissen}, \citenamefont {Ness}, \citenamefont {Peterson}, \citenamefont
  {Wilson},\ and\ \citenamefont {Cates}}]{gillissen2019constitutive}%
  \BibitemOpen
  \bibfield  {author} {\bibinfo {author} {\bibfnamefont {J.~J.~J.}\
  \bibnamefont {Gillissen}}, \bibinfo {author} {\bibfnamefont {C.}~\bibnamefont
  {Ness}}, \bibinfo {author} {\bibfnamefont {J.~D.}\ \bibnamefont {Peterson}},
  \bibinfo {author} {\bibfnamefont {H.~J.}\ \bibnamefont {Wilson}}, \ and\
  \bibinfo {author} {\bibfnamefont {M.~E.}\ \bibnamefont {Cates}},\ }\bibfield
  {title} {\enquote {\bibinfo {title} {Constitutive model for time-dependent
  flows of shear-thickening suspensions},}\ }\href@noop {} {\bibfield
  {journal} {\bibinfo  {journal} {Phys. Rev. Lett.}\ }\textbf {\bibinfo
  {volume} {123}},\ \bibinfo {pages} {214504} (\bibinfo {year}
  {2019})}\BibitemShut {NoStop}%
\bibitem [{\citenamefont {Gillissen}\ and\ \citenamefont
  {Wilson}(2018)}]{gillissen2018modeling}%
  \BibitemOpen
  \bibfield  {author} {\bibinfo {author} {\bibfnamefont {J.~J.~J.}\
  \bibnamefont {Gillissen}}\ and\ \bibinfo {author} {\bibfnamefont
  {HJ}~\bibnamefont {Wilson}},\ }\bibfield  {title} {\enquote {\bibinfo {title}
  {Modeling sphere suspension microstructure and stress},}\ }\href@noop {}
  {\bibfield  {journal} {\bibinfo  {journal} {Phys. Rev. E}\ }\textbf {\bibinfo
  {volume} {98}},\ \bibinfo {pages} {033119} (\bibinfo {year}
  {2018})}\BibitemShut {NoStop}%
\bibitem [{\citenamefont {Gadala‐Maria}\ and\ \citenamefont
  {Acrivos}(1980)}]{Gadala-maria1980}%
  \BibitemOpen
  \bibfield  {author} {\bibinfo {author} {\bibfnamefont {F.}~\bibnamefont
  {Gadala‐Maria}}\ and\ \bibinfo {author} {\bibfnamefont {Andreas}\
  \bibnamefont {Acrivos}},\ }\bibfield  {title} {\enquote {\bibinfo {title}
  {Shear‐{Induced} {Structure} in a {Concentrated} {Suspension} of {Solid}
  {Spheres}},}\ }\href {\doibase 10.1122/1.549584} {\bibfield  {journal}
  {\bibinfo  {journal} {J. Rheol.}\ }\textbf {\bibinfo {volume} {24}},\
  \bibinfo {pages} {799--814} (\bibinfo {year} {1980})}\BibitemShut {NoStop}%
\bibitem [{\citenamefont {Ness}\ and\ \citenamefont {Sun}(2016)}]{Ness2016b}%
  \BibitemOpen
  \bibfield  {author} {\bibinfo {author} {\bibfnamefont {Christopher}\
  \bibnamefont {Ness}}\ and\ \bibinfo {author} {\bibfnamefont {Jin}\
  \bibnamefont {Sun}},\ }\bibfield  {title} {\enquote {\bibinfo {title}
  {Two-scale evolution during shear reversal in dense suspensions},}\ }\href
  {\doibase 10.1103/PhysRevE.93.012604} {\bibfield  {journal} {\bibinfo
  {journal} {Phys. Rev. E}\ }\textbf {\bibinfo {volume} {93}},\ \bibinfo
  {pages} {012604} (\bibinfo {year} {2016})}\BibitemShut {NoStop}%
\bibitem [{\citenamefont {Peters}\ \emph {et~al.}(2016)\citenamefont {Peters},
  \citenamefont {Giovanni}, \citenamefont {Gallier}, \citenamefont {Blanc},
  \citenamefont {Lemaire},\ and\ \citenamefont {Lobry}}]{peters_rheology_2016}%
  \BibitemOpen
  \bibfield  {author} {\bibinfo {author} {\bibfnamefont {François}\
  \bibnamefont {Peters}}, \bibinfo {author} {\bibfnamefont {Ghigliotti}\
  \bibnamefont {Giovanni}}, \bibinfo {author} {\bibfnamefont {Stany}\
  \bibnamefont {Gallier}}, \bibinfo {author} {\bibfnamefont {Frédéric}\
  \bibnamefont {Blanc}}, \bibinfo {author} {\bibfnamefont {Elisabeth}\
  \bibnamefont {Lemaire}}, \ and\ \bibinfo {author} {\bibfnamefont {Laurent}\
  \bibnamefont {Lobry}},\ }\bibfield  {title} {\enquote {\bibinfo {title}
  {Rheology of non-{Brownian} suspensions of rough frictional particles under
  shear reversal: {A} numerical study},}\ }\href {\doibase 10.1122/1.4954250}
  {\bibfield  {journal} {\bibinfo  {journal} {J. Rheol.}\ }\textbf {\bibinfo
  {volume} {60}},\ \bibinfo {pages} {715--732} (\bibinfo {year}
  {2016})}\BibitemShut {NoStop}%
\bibitem [{\citenamefont {Goddard}(2006)}]{goddard2006dissipative}%
  \BibitemOpen
  \bibfield  {author} {\bibinfo {author} {\bibfnamefont {JD}~\bibnamefont
  {Goddard}},\ }\bibfield  {title} {\enquote {\bibinfo {title} {A dissipative
  anisotropic fluid model for non-colloidal particle dispersions},}\
  }\href@noop {} {\bibfield  {journal} {\bibinfo  {journal} {J. Fluid Mech.}\
  }\textbf {\bibinfo {volume} {568}},\ \bibinfo {pages} {1--17} (\bibinfo
  {year} {2006})}\BibitemShut {NoStop}%
\bibitem [{\citenamefont {Chacko}\ \emph {et~al.}(2018)\citenamefont {Chacko},
  \citenamefont {Mari}, \citenamefont {Fielding},\ and\ \citenamefont
  {Cates}}]{chacko2018shear}%
  \BibitemOpen
  \bibfield  {author} {\bibinfo {author} {\bibfnamefont {Rahul~N}\ \bibnamefont
  {Chacko}}, \bibinfo {author} {\bibfnamefont {Romain}\ \bibnamefont {Mari}},
  \bibinfo {author} {\bibfnamefont {Suzanne~M}\ \bibnamefont {Fielding}}, \
  and\ \bibinfo {author} {\bibfnamefont {Michael~E}\ \bibnamefont {Cates}},\
  }\bibfield  {title} {\enquote {\bibinfo {title} {Shear reversal in dense
  suspensions: The challenge to fabric evolution models from simulation
  data},}\ }\href@noop {} {\bibfield  {journal} {\bibinfo  {journal} {J. Fluid
  Mech.}\ }\textbf {\bibinfo {volume} {847}},\ \bibinfo {pages} {700--734}
  (\bibinfo {year} {2018})}\BibitemShut {NoStop}%
\bibitem [{\citenamefont {Lin}\ \emph {et~al.}(2016)\citenamefont {Lin},
  \citenamefont {Ness}, \citenamefont {Cates}, \citenamefont {Sun},\ and\
  \citenamefont {Cohen}}]{lin2016tunable}%
  \BibitemOpen
  \bibfield  {author} {\bibinfo {author} {\bibfnamefont {Neil Y.~C.}\
  \bibnamefont {Lin}}, \bibinfo {author} {\bibfnamefont {Christopher}\
  \bibnamefont {Ness}}, \bibinfo {author} {\bibfnamefont {Michael~E}\
  \bibnamefont {Cates}}, \bibinfo {author} {\bibfnamefont {Jin}\ \bibnamefont
  {Sun}}, \ and\ \bibinfo {author} {\bibfnamefont {Itai}\ \bibnamefont
  {Cohen}},\ }\bibfield  {title} {\enquote {\bibinfo {title} {Tunable shear
  thickening in suspensions},}\ }\href@noop {} {\bibfield  {journal} {\bibinfo
  {journal} {P. Natl. A. Sci.}\ }\textbf {\bibinfo {volume} {113}},\ \bibinfo
  {pages} {10774--10778} (\bibinfo {year} {2016})}\BibitemShut {NoStop}%
\bibitem [{\citenamefont {Ness}\ \emph {et~al.}(2018)\citenamefont {Ness},
  \citenamefont {Mari},\ and\ \citenamefont {Cates}}]{ness2018shaken}%
  \BibitemOpen
  \bibfield  {author} {\bibinfo {author} {\bibfnamefont {Christopher}\
  \bibnamefont {Ness}}, \bibinfo {author} {\bibfnamefont {Romain}\ \bibnamefont
  {Mari}}, \ and\ \bibinfo {author} {\bibfnamefont {Michael~E}\ \bibnamefont
  {Cates}},\ }\bibfield  {title} {\enquote {\bibinfo {title} {Shaken and
  stirred: Random organization reduces viscosity and dissipation in granular
  suspensions},}\ }\href@noop {} {\bibfield  {journal} {\bibinfo  {journal}
  {Science Advances}\ }\textbf {\bibinfo {volume} {4}},\ \bibinfo {pages}
  {eaar3296} (\bibinfo {year} {2018})}\BibitemShut {NoStop}%
\bibitem [{\citenamefont {Cates}\ \emph {et~al.}(1998)\citenamefont {Cates},
  \citenamefont {Wittmer}, \citenamefont {Bouchaud},\ and\ \citenamefont
  {Claudin}}]{cates1998jamming}%
  \BibitemOpen
  \bibfield  {author} {\bibinfo {author} {\bibfnamefont {M.~E.}\ \bibnamefont
  {Cates}}, \bibinfo {author} {\bibfnamefont {JP}~\bibnamefont {Wittmer}},
  \bibinfo {author} {\bibfnamefont {J-P}\ \bibnamefont {Bouchaud}}, \ and\
  \bibinfo {author} {\bibfnamefont {Ph}~\bibnamefont {Claudin}},\ }\bibfield
  {title} {\enquote {\bibinfo {title} {Jamming, force chains, and fragile
  matter},}\ }\href@noop {} {\bibfield  {journal} {\bibinfo  {journal} {Phys.
  Rev. Lett.}\ }\textbf {\bibinfo {volume} {81}},\ \bibinfo {pages} {1841}
  (\bibinfo {year} {1998})}\BibitemShut {NoStop}%
\bibitem [{\citenamefont {Sehgal}\ \emph {et~al.}(2019)\citenamefont {Sehgal},
  \citenamefont {Ramaswamy}, \citenamefont {Cohen},\ and\ \citenamefont
  {Kirby}}]{sehgal2019using}%
  \BibitemOpen
  \bibfield  {author} {\bibinfo {author} {\bibfnamefont {Prateek}\ \bibnamefont
  {Sehgal}}, \bibinfo {author} {\bibfnamefont {Meera}\ \bibnamefont
  {Ramaswamy}}, \bibinfo {author} {\bibfnamefont {Itai}\ \bibnamefont {Cohen}},
  \ and\ \bibinfo {author} {\bibfnamefont {Brian~J.}\ \bibnamefont {Kirby}},\
  }\bibfield  {title} {\enquote {\bibinfo {title} {Using acoustic perturbations
  to dynamically tune shear thickening in colloidal suspensions},}\ }\href@noop
  {} {\bibfield  {journal} {\bibinfo  {journal} {Phys. Rev. Lett.}\ }\textbf
  {\bibinfo {volume} {123}},\ \bibinfo {pages} {128001} (\bibinfo {year}
  {2019})}\BibitemShut {NoStop}%
\bibitem [{\citenamefont {Gillissen}\ and\ \citenamefont
  {Wilson}(2019)}]{gillissen2019effect}%
  \BibitemOpen
  \bibfield  {author} {\bibinfo {author} {\bibfnamefont {J.~J.~J.}\
  \bibnamefont {Gillissen}}\ and\ \bibinfo {author} {\bibfnamefont
  {HJ}~\bibnamefont {Wilson}},\ }\bibfield  {title} {\enquote {\bibinfo {title}
  {Effect of normal contact forces on the stress in shear rate invariant
  particle suspensions},}\ }\href@noop {} {\bibfield  {journal} {\bibinfo
  {journal} {Phys. Rev. Fluids}\ }\textbf {\bibinfo {volume} {4}},\ \bibinfo
  {pages} {013301} (\bibinfo {year} {2019})}\BibitemShut {NoStop}%
\bibitem [{\citenamefont {Kim}\ and\ \citenamefont
  {Karrila}(1991)}]{kim1991microhydrodynamics}%
  \BibitemOpen
  \bibfield  {author} {\bibinfo {author} {\bibfnamefont {S}~\bibnamefont
  {Kim}}\ and\ \bibinfo {author} {\bibfnamefont {S}~\bibnamefont {Karrila}},\
  }\href@noop {} {\emph {\bibinfo {title} {Microhydrodynamics: principles and
  selected applications}}}\ (\bibinfo  {publisher} {Butterworth-Heinemann,
  Boston},\ \bibinfo {year} {1991})\BibitemShut {NoStop}%
\bibitem [{\citenamefont {Seto}\ and\ \citenamefont
  {Giusteri}(2018)}]{seto2018normal}%
  \BibitemOpen
  \bibfield  {author} {\bibinfo {author} {\bibfnamefont {Ryohei}\ \bibnamefont
  {Seto}}\ and\ \bibinfo {author} {\bibfnamefont {Giulio~G}\ \bibnamefont
  {Giusteri}},\ }\bibfield  {title} {\enquote {\bibinfo {title} {Normal stress
  differences in dense suspensions},}\ }\href@noop {} {\bibfield  {journal}
  {\bibinfo  {journal} {J Fluid Mech.}\ }\textbf {\bibinfo {volume} {857}},\
  \bibinfo {pages} {200--215} (\bibinfo {year} {2018})}\BibitemShut {NoStop}%
\bibitem [{\citenamefont {Hinch}\ and\ \citenamefont
  {Leal}(1976)}]{hinch1976constitutive}%
  \BibitemOpen
  \bibfield  {author} {\bibinfo {author} {\bibfnamefont {E~J}\ \bibnamefont
  {Hinch}}\ and\ \bibinfo {author} {\bibfnamefont {L~G}\ \bibnamefont {Leal}},\
  }\bibfield  {title} {\enquote {\bibinfo {title} {Constitutive equations in
  suspension mechanics. {P}art 2. {A}pproximate forms for a suspension of rigid
  particles affected by {B}rownian rotations},}\ }\href@noop {} {\bibfield
  {journal} {\bibinfo  {journal} {J. Fluid Mech.}\ }\textbf {\bibinfo {volume}
  {76}},\ \bibinfo {pages} {187--208} (\bibinfo {year} {1976})}\BibitemShut
  {NoStop}%
\bibitem [{\citenamefont {Jeffrey}\ and\ \citenamefont
  {Onishi}(1984)}]{jeffrey1984calculation}%
  \BibitemOpen
  \bibfield  {author} {\bibinfo {author} {\bibfnamefont {DJ}~\bibnamefont
  {Jeffrey}}\ and\ \bibinfo {author} {\bibfnamefont {Y}~\bibnamefont
  {Onishi}},\ }\bibfield  {title} {\enquote {\bibinfo {title} {Calculation of
  the resistance and mobility functions for two unequal rigid spheres in
  low-reynolds-number flow},}\ }\href@noop {} {\bibfield  {journal} {\bibinfo
  {journal} {J. Fluid Mech.}\ }\textbf {\bibinfo {volume} {139}},\ \bibinfo
  {pages} {261--290} (\bibinfo {year} {1984})}\BibitemShut {NoStop}%
\bibitem [{\citenamefont {Jeffrey}(1992)}]{jeffrey1992calculation}%
  \BibitemOpen
  \bibfield  {author} {\bibinfo {author} {\bibfnamefont {DJ}~\bibnamefont
  {Jeffrey}},\ }\bibfield  {title} {\enquote {\bibinfo {title} {The calculation
  of the low reynolds number resistance functions for two unequal spheres},}\
  }\href@noop {} {\bibfield  {journal} {\bibinfo  {journal} {Phys. Fluids}\
  }\textbf {\bibinfo {volume} {4}},\ \bibinfo {pages} {16--29} (\bibinfo {year}
  {1992})}\BibitemShut {NoStop}%
\bibitem [{\citenamefont {Ball}\ and\ \citenamefont
  {Melrose}(1997)}]{ball1997simulation}%
  \BibitemOpen
  \bibfield  {author} {\bibinfo {author} {\bibfnamefont {RC}~\bibnamefont
  {Ball}}\ and\ \bibinfo {author} {\bibfnamefont {John~R}\ \bibnamefont
  {Melrose}},\ }\bibfield  {title} {\enquote {\bibinfo {title} {A simulation
  technique for many spheres in quasi-static motion under frame-invariant pair
  drag and brownian forces},}\ }\href@noop {} {\bibfield  {journal} {\bibinfo
  {journal} {Physica A}\ }\textbf {\bibinfo {volume} {247}},\ \bibinfo {pages}
  {444--472} (\bibinfo {year} {1997})}\BibitemShut {NoStop}%
\bibitem [{\citenamefont {Cheal}\ and\ \citenamefont
  {Ness}(2018)}]{cheal2018rheology}%
  \BibitemOpen
  \bibfield  {author} {\bibinfo {author} {\bibfnamefont {Oliver}\ \bibnamefont
  {Cheal}}\ and\ \bibinfo {author} {\bibfnamefont {Christopher}\ \bibnamefont
  {Ness}},\ }\bibfield  {title} {\enquote {\bibinfo {title} {Rheology of dense
  granular suspensions under extensional flow},}\ }\href@noop {} {\bibfield
  {journal} {\bibinfo  {journal} {J. Rheol.}\ }\textbf {\bibinfo {volume}
  {62}},\ \bibinfo {pages} {501--512} (\bibinfo {year} {2018})}\BibitemShut
  {NoStop}%
\bibitem [{\citenamefont {Cundall}\ and\ \citenamefont
  {Strack}(1979)}]{cundall1979discrete}%
  \BibitemOpen
  \bibfield  {author} {\bibinfo {author} {\bibfnamefont {Peter~A}\ \bibnamefont
  {Cundall}}\ and\ \bibinfo {author} {\bibfnamefont {Otto~DL}\ \bibnamefont
  {Strack}},\ }\bibfield  {title} {\enquote {\bibinfo {title} {A discrete
  numerical model for granular assemblies},}\ }\href@noop {} {\bibfield
  {journal} {\bibinfo  {journal} {Geotechnique}\ }\textbf {\bibinfo {volume}
  {29}},\ \bibinfo {pages} {47--65} (\bibinfo {year} {1979})}\BibitemShut
  {NoStop}%
\bibitem [{\citenamefont {Plimpton}(1995)}]{plimpton1995fast}%
  \BibitemOpen
  \bibfield  {author} {\bibinfo {author} {\bibfnamefont {Steve}\ \bibnamefont
  {Plimpton}},\ }\bibfield  {title} {\enquote {\bibinfo {title} {Fast parallel
  algorithms for short-range molecular dynamics},}\ }\href@noop {} {\bibfield
  {journal} {\bibinfo  {journal} {J. Comput. Phys.}\ }\textbf {\bibinfo
  {volume} {117}},\ \bibinfo {pages} {1--19} (\bibinfo {year}
  {1995})}\BibitemShut {NoStop}%
\end{thebibliography}%

\end{document}